\documentclass[aps,pra,twocolumn]{revtex4-2}
\usepackage{babel}
   
\usepackage[utf8]{inputenc} % allow utf-8 input
\usepackage[T1]{fontenc}    % use 8-bit T1 fonts
\usepackage{url}            % simple URL typesetting
\usepackage{booktabs}       % professional-quality tables
\usepackage{amsfonts}       % blackboard math symbols
\usepackage{braket} 
\usepackage{varioref}
\usepackage{hyperref}

\usepackage{amsmath} 
\usepackage{nameref}
\usepackage{graphicx}       % tools for inserting graphics
\usepackage[ruled,vlined]{algorithm2e}  % pseudocode insertion
\usepackage{enumitem}       % allow to modify the layout for list environments

\usepackage{amsthm}         % theorem setup
\usepackage{xspace}         % smart spaces after a command
\usepackage{subfig}         % several subfigures in one figure
\usepackage{bbold}          % for the characteristic function
\usepackage{qcircuit}       % quantum circuit drawing
\usepackage{float}          % figures placement
\usepackage{svg}            % figure in svg format
\usepackage{array}
\usepackage{mathtools}
\usepackage{physics}
\usepackage{wrapfig}
\usepackage{graphicx}
\usepackage{tikz}
\usepackage{tabularx}
\usepackage{bbold}
\mathtoolsset{showonlyrefs=true}

\newtheorem{theorem}{Theorem}[section]
\newtheorem{corollary}{Corollary}[theorem]

\newcommand\underrel[3][]{\mathrel{\mathop{#3}\limits_{%
      \ifx c#1\relax\mathclap{#2}\else#2\fi}}}
      
\renewcommand{\Re}{\operatorname{Re}}
\renewcommand{\Im}{\operatorname{Im}}
\renewcommand{\hat}{}

\definecolor{ao(english)}{rgb}{0.0, 0.5, 0.0}

\begin{document}

\title{Twisted quantum walks, generalised Dirac equation and Fermion doubling}

\author{Nicolas Jolly}
\affiliation{Ecole Normale Supérieure de Lyon, France \\ \& MINES Paris, Université PSL, France \\ \& Aix-Marseille Université, Université de Toulon, CNRS, LIS, Marseille, France}

\author{Giuseppe Di Molfetta\footnote{Corresponding author.}}
\email{giuseppe.dimolfetta@lis-lab.fr}
\affiliation{Aix-Marseille Université, Université de Toulon, CNRS, LIS, Marseille, France}

\date{\today}
\begin{abstract}
\begin{center}
\bf{Abstract}
\end{center}
Quantum discrete-time walkers have since their introduction demonstrated applications in algorithmics and to model and simulate a wide range of transport phenomena. They have long been considered the discrete-time and discrete space analogue of the Dirac equation and have been used as a primitive to simulate quantum field theories precisely because of some of their internal symmetries. In this paper we introduce a new family of quantum walks, said \textit{twisted}, which admits, as continuous limit, a generalised Dirac operator equipped with a dispersion term. Moreover, this quadratic term in the energy spectrum acts as an effective mass, leading to a regularization of the well known Fermion doubling problem.  
\end{abstract}
%\pacs{}
\keywords{Quantum Walks}
\maketitle

\section{Introduction}

Quantum walks have occupied a prominent role in recent years for both their algorithmic applications and their ability to simulate physical phenomena. Made popular by a renowned 1993 paper by Aharonov, Davidovich, and Zagury~\cite{aharonov1993quantum}, they had already been studied by Grossing and \textcolor{black}{Zeilinger} as the one-particle sector of a quantum cellular automata in 1988~\cite{grossing1988quantum}. Their realisation may come in two different fashions: in continuous time and in discrete time. Historically, each has followed an independent path, often having different fields of application~\cite{venegas2012quantum}. The former has been more successful in establishing itself in algorithmic search and optimisation applications, but also appear in graph isomorphism problems \cite{gamble2010two}, \textcolor{black}{ranking nodes in a network \cite{boito2022ranking} and pattern-recognition \cite{rossi2013continuous}}. Discrete-time quantum walkers have had the widest application in the quantum simulation of transport phenomena. The main reason for this is threefold. The necessary presence of a space internal to the walker, called a quantum coin, allows a very broad spectrum of physical phenomena to be simulated~\cite{arnault2020quantum, di2013quantum, di2014quantum, di2016quantum, arnault2016quantum, marquez2018electromagnetic, marquez2017fermion, hatifi2019quantum}. Many families of quantum walkers have been proven to retain many of the internal physical symmetries that are often violated by the usual methods of discretization~\cite{arrighi2014discrete, debbasch2019discrete, di2012discrete}. The natural multi-particle extension, namely quantum cellular automata, are found to be easily physically implementable on current quantum platforms, such as those based on the Rydberg atom~\cite{henriet2020quantum, wintermantel2020unitary, sellapillay2022entanglement}. 
Although fundamentally different, quantum walkers in continuous time and discrete time have been compared for a long time~\cite{strauch2006connecting, childs2010relationship}. The motivation was essentially to understand whether the former could be the continuous-time limit of the latter. \textcolor{black}{Although, Child et al. \cite{childs2010relationship} proved how to simulate the dynamics of a continuous-time quantum walk on any graph using discrete-time quantum walks, this proof did not account the internal coin state space of the walker in the limit procedure. Only recently, such a proof has been extended to a braod class of coined quantium walks, leading, in the continuous limit to the Dirac equation. Such new family of quantum walks has been called \textit{plastic quantum walkers} (PQW)~\cite{di2020quantum,manighalam2021continuous},  which admit both the continuous limit in time and the continuous limit in space-time. Moreover, this result, for the first time, proved possible to use the same circuit to simulate the Dirac equation and the corresponding Hamiltonian dynamics on a lattice in continuous time.}

This plasticity property has been rigorously investigated in one and two spatial dimensions. In both cases, it has been proven that the only possible non-trivial Hamiltonian obtainable with a plastic quantum walker is the Dirac one. This is not surprising, considering that quantum walk in discrete time have been for a long time considered the discrete analogue of the Dirac equation~\cite{arrighi2014dirac}. As byproduct, the non-relativistic limit of the quantum walks equations has been always recovered by non-unitary operations. 

In this manuscript, we introduce a more general family of plastic quantum walk. These, which we call, Twisted Quantum Walks (TQW), while recovering PQWs in a certain limit, allow us to simulate both the Dirac operator as well as the Laplacian operator at first order. This result has multiple consequences: overall it contributes to making the spectrum of phenomena that a quantum walk can simulate even wider. These in fact offer a simulation protocol for two of the most important equations in physics, the Dirac equation and the Schrödinger equation. Furthermore, since the Laplacian and the Dirac operator are of the same order, our scheme converges to an equation that curiously resembles the Fokker-Plank equation in imaginary time or to a second order Dirac equation, known to model graphene electrons in an electromagnetic field~\cite{falomir2018optical}.
Moreover, quite unexpectedly, we have shown that such QW families are inherently free of the Fermion doubling problem, offering themselves as primitives for constructing lattice field theories. At the same time, we take advantage of the entire theoretical framework on quantum walks and their generalisation to multiple particles, the quantum cellular automata. 

It only remains to anticipate why we have named this family of PQWs, twisted. As is well known, a quantum walker is a sequence of two unitary operators. The first one acts in the inner space of the walker, the quantum coin. The second is a motion operator conditioned by the internal state of the walker. A plastic QW is a walker with a tunable speed of propagation. Indeed intuitively, during the process where the continous-time limit discrete-space is taken, whenever $\Delta_t$ gets halved relative to $\Delta_x$, so is the particle’s speed, because it gets half the time to propagate. This plasticity, achieved with a suitable anisotropic coin parameterisation, allows the Dirac equation to be simulated in continuous space-time as well as the Dirac Hamiltonian on a lattice in continuous time. 

In Twisted QWs, we introduce a new operation in the definition of PQWs, namely a rotation in the coin basis for some and not for all walker moves. This twist in the walker's inner space dramatically affects the dynamics, as quantum coin states are known to be intricately entangled with position states. The result is a more general QW, which converges in general to a generalised Dirac equation, including dispersive terms. Moreover, some twists also introduce a chirality breaking in the QW operator. A well know sufficient condition to avoid the Fermion doubling problem~\cite{nielsen1981no}. 

The manuscript is organised as follows: Sec.\ref{sec:themodel} is devoted to introduce the general model of the twisted quantum walk, while in Sec.\ref{sec:app1} and Sec.\ref{sec:app2} we study respectively an homogeneous and non-homogeneous twist. Both cases are carefully investigated from a qualitative and theoretical point of view and for each case we compute formally the continuous equations. Finally in Sec.~\ref{sec:conc} we conclude. The manuscript includes a detailed appendix in Sec.~\ref{sec:annex}, where we provide a rigorous proof for the continuous limit and further considerations about the phenomenology of the TQW.  

\section{The twisted quantum walk}
\label{sec:themodel}

A discrete time quantum walk on a line is the quantum analog of a one-dimensional random walk. It lives in a composite Hilbert space: the coin state space, spanned by the orthonormal basis $\{\ket{+},\ket{-}\}$, encoding the walker direction and the position state space. The position is labeled by $\ket{l}\in\mathbb{Z}$. Then, a generic state of the walker at the instant $j$ reads as follows:
\begin{small}
\begin{equation}
\Psi_j= \sum_{v\in\{+,-\}}\sum_{l\in \mathbb{Z}} \psi^v_{j,l} \ket{v,l},
\end{equation}
\end{small}
 We introduce the discretization steps in time and space,  respectively $\Delta_t$ and $\Delta_x$, such that $\psi^v_{j,l}$ approximates a continuous function $\tilde{\psi}^v(t, x)$ defined on $\mathbb{R^+} \times \mathbb{R}$. The two functions match on the vertices of the grid by $\psi^v_{j,l}= \tilde{\psi}^v(t= j\Delta_t, x=l\Delta_x)$. In this paper, we will refer only to $\psi^v(t,x)$ for more clarity. The time evolution of a quantum walk is driven by a coin operator which acts in the internal coin-state space plus a conditional shift moving forward the $\ket{+}$ component and backward the $\ket{-}$ component. The operator $S_x$ is the usual state-dependent shift operator $\hat{S}_x \ket{v}\ket{i} = \ket{v}\ket{i-(-1)^v}$. The coin operators $C$ are in general elements of $U(2)$ and are defined by four real parameters $(\delta, \theta, \phi, \zeta)$. The most general expression is:
\begin{equation}
\begin{aligned}
    C(\delta, \theta, \phi, \zeta ) &:= e^{i \delta} R_{z}\left(\zeta\right) R_{y}\left(\theta\right) R_{z}\left(\phi\right) \\
   % &= e^{i \delta} e^{-i \zeta \sigma_{z} / 2} e^{-i \theta \sigma_{y} / 2} e^{-i \phi_{j} \sigma_{z} / 2} \\
    &= e^{i \delta}\left(\begin{array}{l}
\cos \frac{\theta}{2} \exp -i \frac{\phi+\zeta}{2}-\sin \frac{\theta}{2} \exp i \frac{\phi-\zeta}{2} \\  \\
\sin \frac{\theta}{2} \exp i \frac{-\phi+\zeta}{2} \quad \cos \frac{\theta}{2} \exp i \frac{\phi+\zeta}{2}
\end{array}\right).
    \end{aligned}
\end{equation}

The novelty expressed in this manuscript is to introduce an extra operation, $T$, namely a rotation (\textit{aka} twist) from the computational basis $\{\ket{v^-}, \ket{v^+}\}$ to an arbitrary one. The overall twisted dynamics obeys to the following equation:
\begin{equation}
\Psi(t+\Delta_t)= U_{T} \Psi(t):= M G_T \Psi(t)
\label{eq:FDE0}
\end{equation}
where
\begin{equation}
\begin{aligned}
M &:= ~ \exp (i \mu \sigma_z)\\
G &:= ~W_{\beta} W_{\alpha},
\hspace{0.3cm}\text{with:}\hspace{0.2cm} &  \left\{ 
\begin{aligned} W_{\alpha}:=& ~S_x C_{\alpha} \\
W_{\beta}:=& ~T^{-1} S_x T C_{\beta} .
\end{aligned} \right.
\end{aligned}
\end{equation}
The operator $M$ has an importance for covering the mass term in the propagation equations in the continuous limit. Included here to be as general as possible, it does not play any fundamental role in the following sections. Notice that the $W_{\alpha}$ coincides with the usual quantum walk operator.
In what follows, we will study the above twisted QW (TQW) in two specific cases, leaving a more general discussion to a sequel. In the first specific case the twist $T$ is homogeneous, whereas in the second case the twists are two different ones and are alternated all along the evolution. Both cases will be studied from several angles: first a qualitative look at the spectrum and the dynamic behaviour of the walker. Then a rigorous proof of the continuous limit will be provided.

\textcolor{black}{It should also be noted that the twist operation cannot be assimilated to a reparametrization of the coin. In fact, one of the authors rigorously proved that no family of QW in 1+1d and 1+2d admits second-order terms in the spatial derivative at the continuous limit~\cite{di2020quantum,manighalam2021continuous}. The proof is in fact general and is valid for any choice of coin operator and any stroboscopic period for which the continuous limit is calculated. The twist also coincides, as can be seen from equation~\eqref{eq:FDE0}, with a rotation of the base of the displacement operator, and not of the coin operator. This is reminiscent of the base change required when the walker changes direction in a d-dimensional grid.} 

%, with the shape $S_x C_{\alpha}$. In the $W_{\beta}$, instead of taking a shift in the $x$ direction as we usually would, we operate a change of base by the matrix $T$ before and after the shift in $x$. We will see through the following examples that this will lead to new behaviours not observed in the usual QW. The changes of base are operators of $SU(2)$ and are generally noted $T$.
%In the $W_2$, we try to mimic the 2D case with the shape $S_y C_2$. Instead of taking a shift in the $y$ direction as we would in 2D, we operate a change of base by the matrix $T$ before and after the shift in $x$. We will see through the following examples that this will lead to new behaviours not observed in the usual QW. The changes of base are operators of $SU(2)$ and are generally noted $T$. 

\section{Homogeneous \textit{Y-Y} twist and the dispersion emergence}
\label{sec:app1}

%\subsection{Model}
%\label{subsec:app1_model}

Throughout this section we will consider the time evolution:
\begin{equation}
\Psi(t+\Delta_t)= U_{YY} \Psi(t)= M G^2_{Y}\Psi(t)
\label{eq:master_eq_W_homo}
\end{equation}
% where the unitary operator $G$ is defined as:
% \begin{equation}
% \begin{aligned}
% U_{YY} &:= ~ (W_{\beta} W_{\alpha})^2 
% \text{with: \quad}&  \left\{ 
% \begin{aligned} W_{\alpha}:=& ~S_x C_{\alpha} \\
% W_{\beta}:=& ~T^{-1} S_x T C_{\beta} \\
% %M:=& ~ \exp (i \mu \sigma_z)
% \end{aligned} \right.
% \end{aligned}
% \end{equation}
For simplicity let us considering the following parameterisation: 
\begin{equation}
\begin{aligned}
C_{\alpha} &:= C(\frac{\pi}{2}, -\pi - 2 \alpha,\frac{\pi}{2}, -\frac{3\pi}{2})=   \begin{pmatrix*}  \sin\alpha & -i \cos \alpha  \\
i \cos \alpha  & -\sin\alpha 
\end{pmatrix*} \\
&= \sigma_y R_x(2\alpha) \\
C_{\beta} &:= C(0, -2\pi - 2 \alpha, \frac{\pi}{2}, -\frac{\pi}{2})=   \begin{pmatrix*}  -\cos \alpha & -i \sin \alpha  \\
-i \sin \alpha  & -\cos \alpha 
\end{pmatrix*} \\
&= - R_x(-2\alpha) \\
T~ &:
= R_y(\theta) 
\end{aligned}    
\end{equation}

\begin{figure}[h]
%\vspace{-5pt}
  %  \setlength{\belowcaptionskip}{-10pt}
   \includegraphics[width=.4\textwidth]{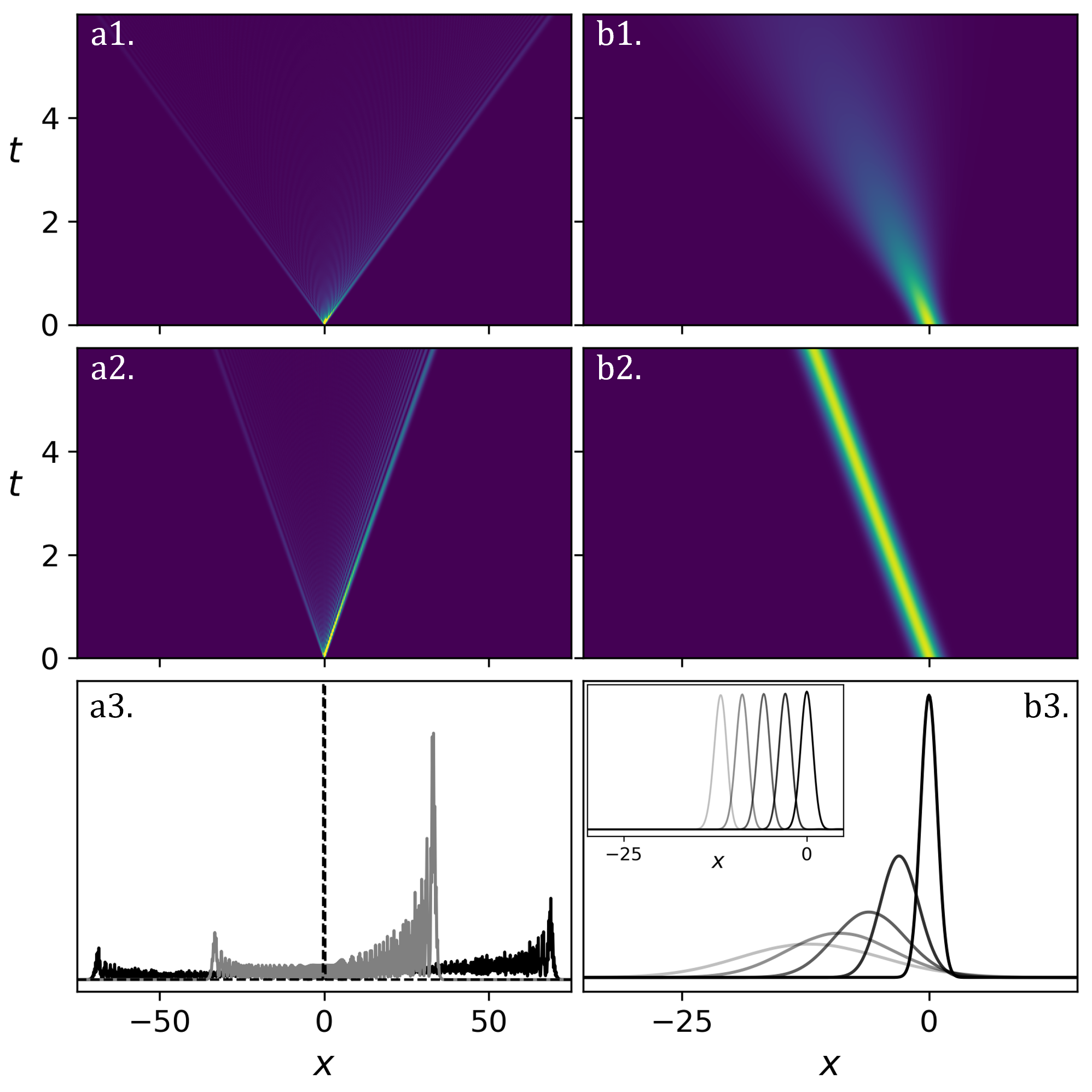}
  % \vspace{-5pt}
   \caption{\raggedright \footnotesize \textcolor{black}{These figures show the time evolution of the system. In the first column \textbf{(a)}, the initial condition is a narrow Gaussian distribution: $\sqrt{g(x)}\otimes (1, 1-i)^{\text{T}}$ with $\sigma^2=0.01$ and $\alpha_1=3$. In the second column \textbf{(b)}, the initial condition is a wide Gaussian: $\sqrt{g(x)}\otimes (1,i)^{\text{T}}$ with $\sigma^2=0.7$, $\alpha_1=1$. In all figures $\varepsilon=0.01$. The first four sub-figures from the top show the density as function of space and time. The lighter the color the higher the density of probability. In the sub-figures a1. and b1. $\theta=\pi/2$, i.e. there is a twist. For sub-figures a2. and b2., $\theta=0$, i.e. no twist. The sub-figure a3. shows the initial distribution of the column in dashed line. It also shows the probability distribution at $t=300$ respectively with ($\theta= \pi/2$) in black and without twist ($\theta= 0$) in gray. The sub-figure b3. shows the probability distribution at several times for $\theta= \pi/2$ and in the inset the same for $\theta= 0$. The lighter the curve, the more advanced in the simulation. }
   }
\label{fig:density_plot_homo}
\end{figure}

Note that the coin operators are non-commutative and the twist operator coincides with a rotation of a $\theta$-angle around $y$. \textcolor{black}{Note that one complete step of the walker corresponds to iterate the operator $G$ twice, with the same twist at each iteration. This choice is justified by the existence conditions of the continuous limit. In fact, as it is rigorously proved in the annex \ref{sec:annex}, the operator G at zero order of development does not coincide with the identity, but only its square does.} In order to have a qualitative idea of the dynamics, we run few numerical simulations playing with the main parameters of the walker. Let us choose as initial state of the walker a Gaussian packet $\psi(x,0)=\sqrt{\rho_0(x, t)}$, with $\rho_0(x)= \frac{1}{\sigma \sqrt{2 \pi}} e^{-\frac{1}{2}\left(\frac{x-\mu}{\sigma}\right)^{2}}$. We can see that the TQW in Fig.~\ref{fig:density_plot_homo}.b clearly displays a pronounced dispersion of the wave packet, with respect to the twist-less case shown in Fig.~\ref{fig:density_plot_homo}.c and in Fig.~\ref{fig:density_plot_homo}.d. As we will see, this dispersion can be fully understood looking at the energy spectrum associated to the effective Hamiltonian of the walker, $H_{\text{eff}}= i \log U_{YY}/ \Delta_t$. The eigenvalues of $H_{\text{eff}}$,  $\lambda_{\text{eff}}$, can be then expressed from the eigenvalues of the unitary operator $U_{YY}$, $\nu$, as $\lambda_{\text{eff}~j}= i \log \nu_j/2\Delta_t \quad \text{ with $j= 1,2$}$.
 
Now the eigenvalues of $U_{YY}$ can be expressed as roots of the characteristic polynomial: 
\begin{equation}
    \nu= \frac{\Tr U_{YY}}{2} \pm \sqrt{ \left (\frac{\Tr U_{YY}}{2} \right) ^2 - \det U_{YY}}= d_0 \pm \sqrt{d_0^2 - 1}
\end{equation}

\begin{figure}[h]
%\vspace{-10pt}
    %\setlength{\belowcaptionskip}{-30pt}
   \includegraphics[width=.48\textwidth]{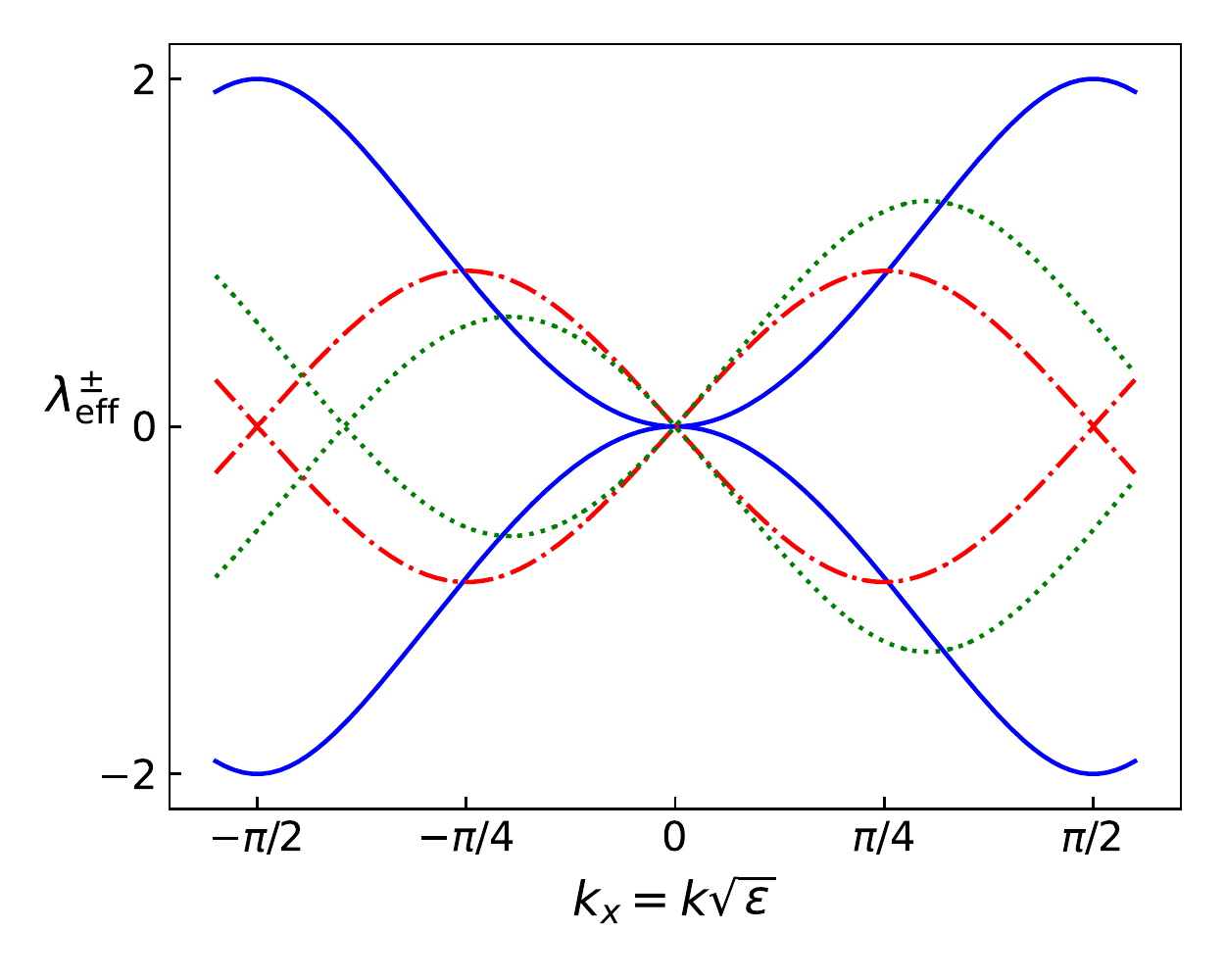}
  % \vspace{-15pt}
   \caption{\footnotesize Effective spectrum of the walk as a function of $k$ for different set of parameters $(\alpha, \theta)$. Solid blue: $(0~,~1)$; Dash-dot red: $(\pi/3~,~ 0)$; Dotted green: $(\pi/3~,~ \pi/5)$. All curves are for the step $\varepsilon= 1$.}
\label{fig:spectrum_eff_homo}
\end{figure}

As $U_{YY}$ is the product of rotations, we have $\det U_{YY}= 1$. We also set $d_0:= \Tr U_{YY}/2 \in \mathbb{R}$ the $\sigma_0$ component of $U_{YY}$ in its decomposition in the Pauli basis of Hermitian $2\cross 2$ matrices. Furthermore, as $W$ is unitary, the eigenvalues are on the unit circle which means that $\log \nu= \text{Arg } \nu$ and $\cos \big[ \text{Arg } \nu \big]= d_0$. In other words, we have:
\begin{equation}
    \lambda_{\text{eff}}= \pm \dfrac{ \arccos d_0 }{2\Delta_t}.
\end{equation}
Hence, the spectrum of the walk is completely determined by $d_0$, which means that we can deduce the spectrum's properties from the solely knowledge of $d_0$. With our choice of parameters and setting here $m=0$, we have: 
\begin{equation}
\begin{aligned}
    d_0 &=  \frac{11-A}{16} + \cos 2 k \cdot \frac{1+A}{4} \\
    &+ \cos 4 k \cdot  \frac{1-3A}{16} \\
    &+ \sin^4 k \cdot \left(\frac{1+A}{2}\right) \left(1-2B^2\right) \\
    &+ \sin^3 k \cos k \cdot 2B \sqrt{1-A^2} \\ \\
    &\text{with: } A:= \cos 4 \alpha \\
    &\text{and: } B:= \sin \theta \\ 
\end{aligned}
\label{eq:d0_homo}
\end{equation}
From this expression, and seeing that $\sin^4 u= \frac{1}{8}(3 - 4 \cos 2u + \cos 4u)$ and $\sin^3 u \cos u= \frac{1}{8}(2 \sin 2u - \sin 4u)$~; we can deduce that the Brillouin Zone is $k \in [-\pi/2, \pi/2]$. We also notice that the only non $k$-symmetric part in $d_0$ is the last term of the above sum, which vanishes under $\theta= n \pi$ (i.e. trivial twists) or $\alpha= n \frac{\pi}{4}$, $n \in \mathbb{Z}$. This asymmetry in the spectrum is unexpected and is a direct consequence of the twist. As we shall see, this results in a drift of the first moment of the distribution. It is particularly interesting that there are non-trivial values of the twist so that the spectrum remains symmetrical. The spectrum of the walk is shown in Fig.~(\ref{fig:spectrum_eff_homo}) for several sets of parameters. 
The red and blue curves are $k$-symmetric because $\sin 4 \alpha \cdot \sin \theta= 0$. The green curve is remarkably not symmetric. We also notice that the curves with $\alpha \neq 0$ all go to zero more than once in the Brillouin Zone. This reminds us the so called Fermion doubling problem. We will see in the following section how to tackle this issue, playing with the twist.

Now, in order to compute the continuous limit, let us introduce the following parameterisation, reminiscent of those introduced by Di Molfetta, Arrighi and Manighalam in~\cite{manighalam2021continuous, di2020quantum}, for the plastic quantum walk:
\begin{equation}
\begin{aligned}
\Delta_t &= \varepsilon \\
\Delta ~ &= \sqrt{\varepsilon} \\ 
k ~ &= k \sqrt{\varepsilon} \\ \\
\alpha ~ &= \sqrt{\varepsilon} \alpha_1 \\
\mu ~ &= m \Delta_t= m \varepsilon
\end{aligned}    
\end{equation}
Note that although these scaling laws appear to treat time and space anisotropically, the spatial derivatives $\alpha \partial_x$ are always of order $O(\varepsilon)$. In fact, the spacetime isotropy is recovered by considering the discrete $x$-axis renormalised by a factor $\alpha^{-1}$. This change of coordinates, parameterised by $\varepsilon$, guarantees plasticity. This property is in fact a \textit{necessary condition} for the twist to have an effect in the leading orders of the Taylor development of the $U$ operator. Moreover, we can remark that we do not need that the angle of the twist scale with $\varepsilon$. 
Then, taking the Taylor expansion of the Eq.~\eqref{eq:d0_homo}, computing the formal limit for $\varepsilon \to 0$ and leaving all the details to the Annex~\ref{subsec:contLim}, we recover: 
\begin{equation}
\begin{aligned}
     & d_0\underrel{\varepsilon \to 0}{\approx} 1 - 2 k^2 (2 \alpha_1 - k \sin \theta)^2 \varepsilon^2 \\
    & \text{plus: } \quad \arccos (1-u)  \underrel{u \underrel{>}{\to} 0}{\approx} \sqrt{2u} \\ \\
    &\text{hence: } \lambda_{\text{cont}}= \pm  k (2\alpha_1 - k \sin \theta).
\end{aligned}
\label{eq:spectrum_cont_homo}
\end{equation}
Thus the whole Hamiltonian $\mathcal{H}_c$ reads:
\begin{equation}
\begin{aligned}
    \mathcal{H}_c &= \begin{pmatrix*}  -\frac{m}{2} & -i k (-2 \alpha_1 + k \sin \theta )\\
+ i k (-2 \alpha_1 + k \sin \theta   )& + \frac{m}{2}
\end{pmatrix*}\\
&= -\dfrac{m}{2} \sigma_z + k(-2 \alpha_1 + k \sin \theta ) \sigma_y
\end{aligned}
\end{equation}
leading to an overall continuous dynamics:
\begin{equation}
    i \partial_t \Psi= \mathcal{H}_c \Psi.
\label{eq:master_eq_homo}
\end{equation}

\begin{figure}[hbtp]
\vspace{-10pt}
   \includegraphics[width=.499\textwidth]{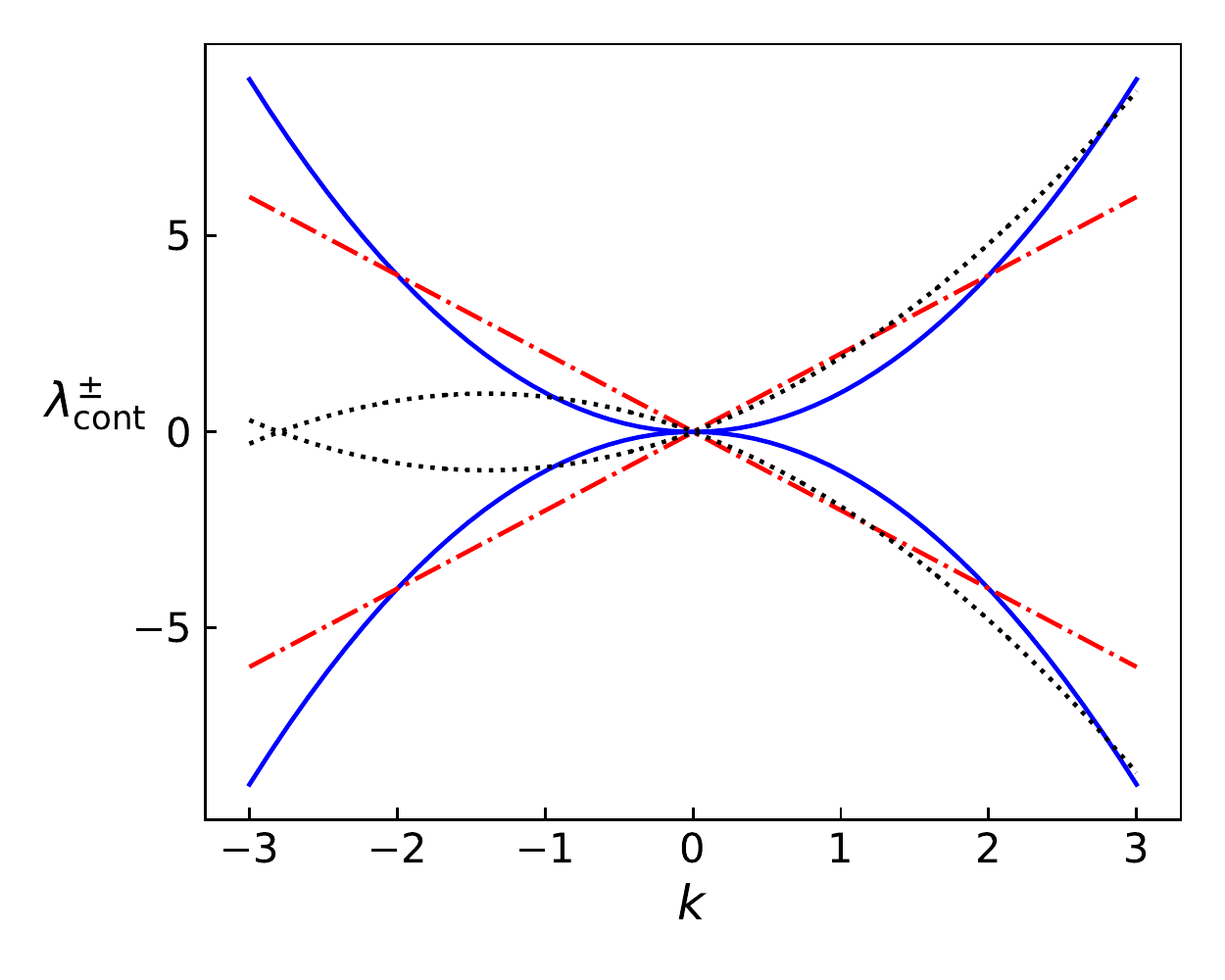}
   \vspace{-15pt}
   \caption{\footnotesize Energy spectrum of the walk as a function of $k$ for different set of parameters $(\alpha_1, \theta)$, in the continuous limit regime. Solid blue: $(0~,~ \pi/2)$; Dash-dot red: $(1~,~ 0)$ ; Dotted green: $(0.7~,~ \pi/6)$. All curves are for $m=0$. We notice that as predicted, when $\theta=0$ (dash-dot red) then the spectrum is linear, i.e. \textit{Dirac-like}, when $\alpha_1= 0$ (solid blue) then the spectrum is quadratic i.e. \textit{Schrödinger-like}. And in general it is in between the two, and asymmetric. }
\label{fig:spectrum_cont_homo}
\end{figure}

From the above, we can conclude that the walk we introduced leads in general to a quadratic spectrum in which the $\alpha_1$ coefficient controls the linear part and the $\theta$ term controls the quadratic part. For $\theta= 0$, the \textit{twist-less} case, we recover the usual (1+1)D Dirac equation, with its linear dispersion relation similar to $E=\pm  \hbar k$. For $\alpha_1= 0$, we have the non relativistic equivalent: the massive Schrödinger equation. For other sets of parameters, the system should be in between the two as shown in Fig.~\ref{fig:spectrum_cont_homo} and can be seen as a second-order Dirac equation or as a Schrödinger equation with a velocity-dependent potential. 

The emergence of the second order derivative at the order $\varepsilon$ is due to the twist $T$. More precisely it allows the off-diagonal terms in $k$ of order $\sqrt{\varepsilon}$ to cancel-out and the $k^2$ terms, scaling as $\sqrt{\varepsilon}^2$, to survive as leading order. Finally, notice that, if we chose to scale the angle of the twist with $\varepsilon$, the $k^2 \sin \theta$ would vanish when taking the continuous limit. In other words, we need to keep the angle of the twist constant, to allow the emergence of the second order derivative. It is worth noting that, the above PDEs confirm that the energy spectrum is not symmetric in general under the transformation $k \to -k$ as the $(2 \alpha_1 - k \sin \theta)$ factor is not.

%We also note that, again, the eigenvectors in the continuous limit and in the massless case are $(  1 , i)^{\text{T}} ; (1 , -i )^{\text{T}}$, which means that for a state initialized in either of these spins, the propagation will be approximately in a fixed direction and without any interference.

\begin{figure}[hbtp]
\vspace{-5pt}
    \setlength{\belowcaptionskip}{-0pt}
   \includegraphics[width=.5\textwidth]{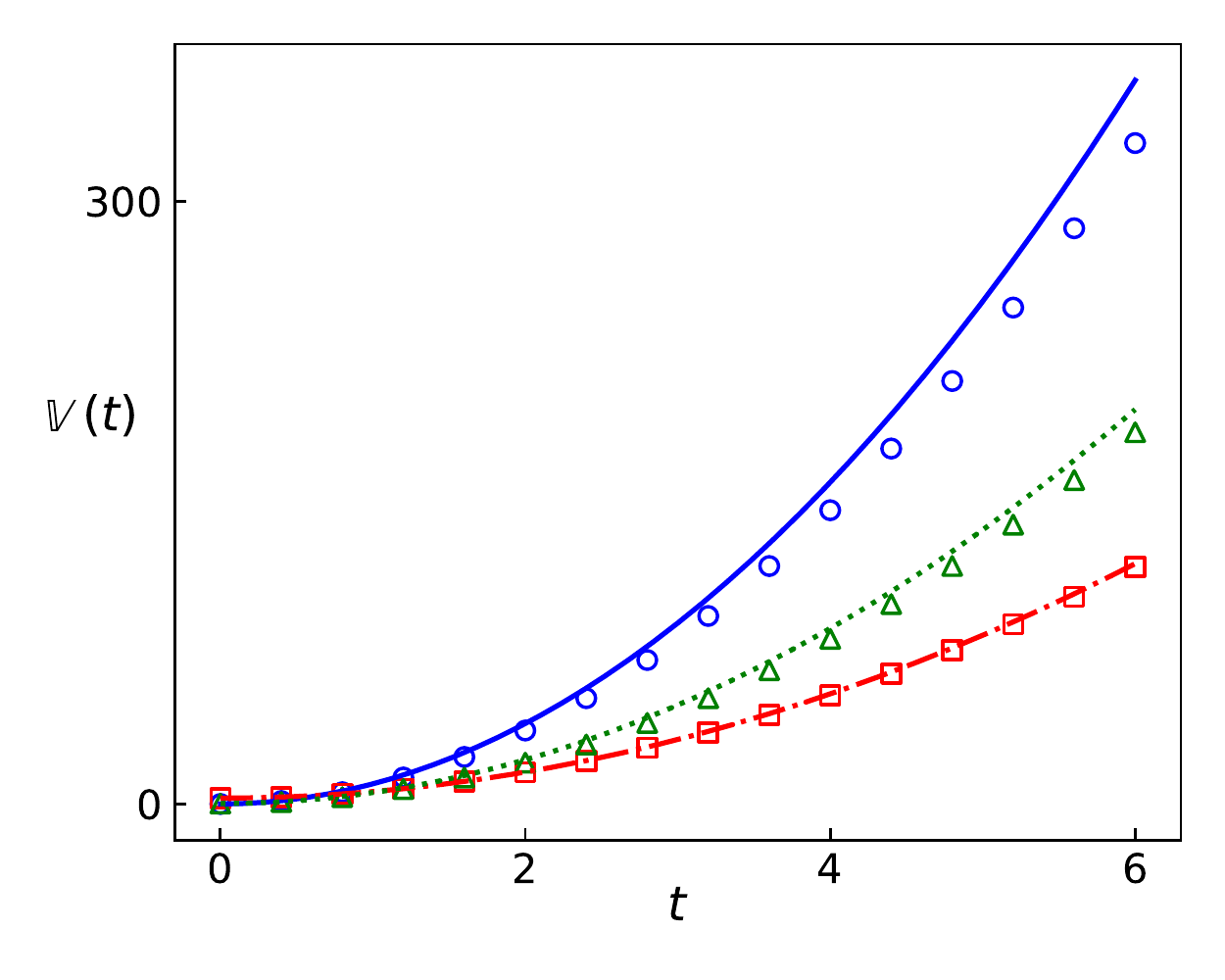}
   \vspace{-15pt}
   \caption{\footnotesize \textcolor{black}{Time evolution of $\mathbb{V}$ for different set of parameters. Data from numerical simulations are represented by dots and the theoretical model following expression (\eqref{eq:var_homo}) in solid lines. The values for the parameters are the following: \textcolor{blue}{$\boldsymbol{\circ}$}:~$\alpha_1= 0~;~\theta= \pi/2~;~\Psi_0 \propto (1,i)^{\text{T}}~;~\sigma= 0.1$. \textcolor{black}{$\boldsymbol{\square}$}:~$\alpha_1= 0.9~;~\theta= 0~;~\Psi_0 \propto (1,0)^{\text{T}}~;~\sigma= 3$. \textcolor{ao(english)}{$\boldsymbol{\triangle}$}:~$\alpha_1= 1.1~;~\theta= 2~;~\Psi_0 \propto (1,1+i)^{\text{T}}~;~\sigma= 0.3$.
     For all data, $\mu=0$, $\varepsilon= 0.01$ and we took $300$ steps. The relative errors between the theoretical prediction and the simulated dare are estimated at respectively $8.7\%$, ; $1.4\%$, ; $7.1\%$}}
    \label{fig:var_time_homo}
\end{figure}

Moreover we can observe that the PDE~\eqref{eq:master_eq_homo} has formally the same structure of a Fokker-Planck equation but in imaginary time. This can be seen notably in Fig.~\ref{fig:density_plot_homo}.f, where we observe the same drift-diffusion behaviour we would have for the one-dimensional Fokker–Planck equation, where over time the distribution widens due to random impulses. However, despite first appearances, there is no evidence of diffusion here. A fact already clear in the derivation of the Eqs~\eqref{eq:master_eq_homo}, it can be further confirmed by a careful study of the walker variance, as shown in Fig.~\ref{fig:var_time_homo}. Let us start with the first moment of the distribution, which in the continuous limit regime and for the massless evolution leads to:
\begin{equation}
    m_1(t)= \mu + 4t\alpha_1 \Im \big[ \psi_+^0 \overline{\psi_-^0} \big].
\label{eq:m1}
\end{equation}
The above equation confirms that the center of the distribution follows a linear drift in time proportional to $\alpha_1$, as it could be seen in the spectrum. In fact, this is consistent with our comparison with the Fokker-Planck Equation. We note that the parameter $\alpha_1$ fully controls the drift. Now the variance, $\mathbb{V}(t)= m_2(t) - m_1(t)^2$, where $m_2(t)= \langle X^2 \rho \rangle$ is the second order moment of the probability distribution reads:
\begin{equation}
    \mathbb{V}(t)= \sigma^2 + \dfrac{\sin^2 \theta}{\sigma^2} \cdot t^2+ 4 \alpha_1 ^2 \bigg(1  - 4\Im \big[ \psi_+^0 \overline{\psi_-^0} \big]^2~\bigg) \cdot t^2
    \label{eq:var_homo}
\end{equation}
As expected the variance is initially at $\sigma^2$ and grows like $t^2$. This differs from the Fokker-Planck case, where the variance increases in $t$. However, differently from the usual QW, here the spreading of the distribution is also controlled by $\theta$, the twist angle. This makes the prefactor of the variance depending on the initial width of the wave packet $\sigma^2$. In other words, the narrower the initial distribution is, the quicker the distribution will spread in space and conversely. Without twist, this dependence in the initial shape of the distribution would not exist.

If we analyze the $\Im \big[ \psi_+^0 \overline{\psi_-^0} \big]$ term a bit more, we find that it is maximal for the eigenstates, $(1, \pm i)^T$. The maximal value in this case is $\pm 1/2$, leading to a maximal drift. As for the spread, if our initial states coincide with the eigenstates, the last term in the variance vanishes but the twist introduces an additional spread for these states inversely proportional to the initial variance. Moreover, as shown in Fig.~\ref{fig:density_plot_homo}, the final state representation clearly shows that with twist, the distribution of the state clearly disperses. Without twist, the distribution only drifts but does not disperse. In this case the comparison with the drift-diffusion case is relevant. When initialized in the eigenstates, the walker behaves like a distribution undergoing drift controlled by $\alpha_1$ and diffusion controlled by $\theta$ as we interpreted from the resemblance between the Eq.~\eqref{eq:master_eq_homo} and the Fokker-Planck equation. Furthermore, in this case, $m_1$ and $\mathbb{V}$ are completely independent from each other.

% \

% In the simulations, we find that for most values the experimental variance follows quite well the expression we derived (\eqref{eq:var_homo}) as represented in Fig.~\ref{fig:var_time_homo}. The examples displayed fit very well the expression we derived considering the assumptions we made, but there remains some discrepancies after some time. It shows the limit of these calculations, they give a good description of the overall qualitative behaviour of the distribution, but they remain approximative because they were derived taking the continuous limit whereas the simulations use the effective hamiltonian with $\varepsilon > 0$.

%\subsection{Entropy of Entanglement}

\section{Alternate twist and Fermion doubling}
\label{sec:app2}
%\subsection{Model}
%\label{subsec:app2_model}

In this section we introduce two majors modifications: first we allow the twist to change in time, but still keeping the overall operator periodic. Secondly, we make the twist angles scaling with $\varepsilon$. In fact, this choice is a necessary condition to be able to calculate the continuous limit of this walker and avoid any divergence. As example, we use two different twists $T_1= R_x( \theta = \sqrt{\varepsilon} \theta_1)$ and $T_2= \mathbb{1}_2$ alternated as follows:
\begin{equation}
\begin{aligned}
U_{XI} &:= ~M W_{\beta 2} W_{\alpha} W_{\beta 1} W_{\alpha}  
 \end{aligned}
 \end{equation}

The major achievement of this section is to show how a non-homogeneous twist avoids the famous Fermion doubling problem. \textcolor{black}{Let us recall shortly that such problems arises because in the continuous limit we recover $2^d$ different equations for $d$ components of the momentum. In our case, e.g., just two continuous variety around the two poles of the Brillouin Zone, because the momentum $k$ has only one component along $x$. More precisely around $k=0$ and $k=\pm\pi$, the Quantum Walks correctly recover a Fermion dynamics. In free field theory we can always agree to populate only the region near $k=0$ with Fermions. However, in an interacting field theory the modes $k\sim\pm\pi$ may become excited, because the excitation of these modes may not disappear with $\Delta_x$ to zero. There are several ways to solve the problem, and all of them violate one the presuppositions of the Nielsen-Ninomiya theorem~\cite{nielsen1981absence}. In one of them, Wilson and Ginsparg~\cite{ginsparg1982remnant} introduced an effective mass term in the spectrum, which violates chiral symmetry. Doing so the doubling in $\pm\pi$ is removed by a gap. In our model, intuitively, the dispersion term acts as an effective Wilson mass on the degeneracy sites.} This leads to a gap in the spectrum which is controllable by tuning the twist angle. Moreover, in the continuous limit we still recover the Dirac equation, because the dispersion terms is now at higher order in $\varepsilon$. Notice that, differently from most all the others stratagems to discretise the Dirac field equation, avoiding the Fermion doubling, this solution is implemented by means of solely local unitaries. 

\begin{figure}[hbtp]
\vspace{-10pt}
   \includegraphics[width=.499\textwidth]{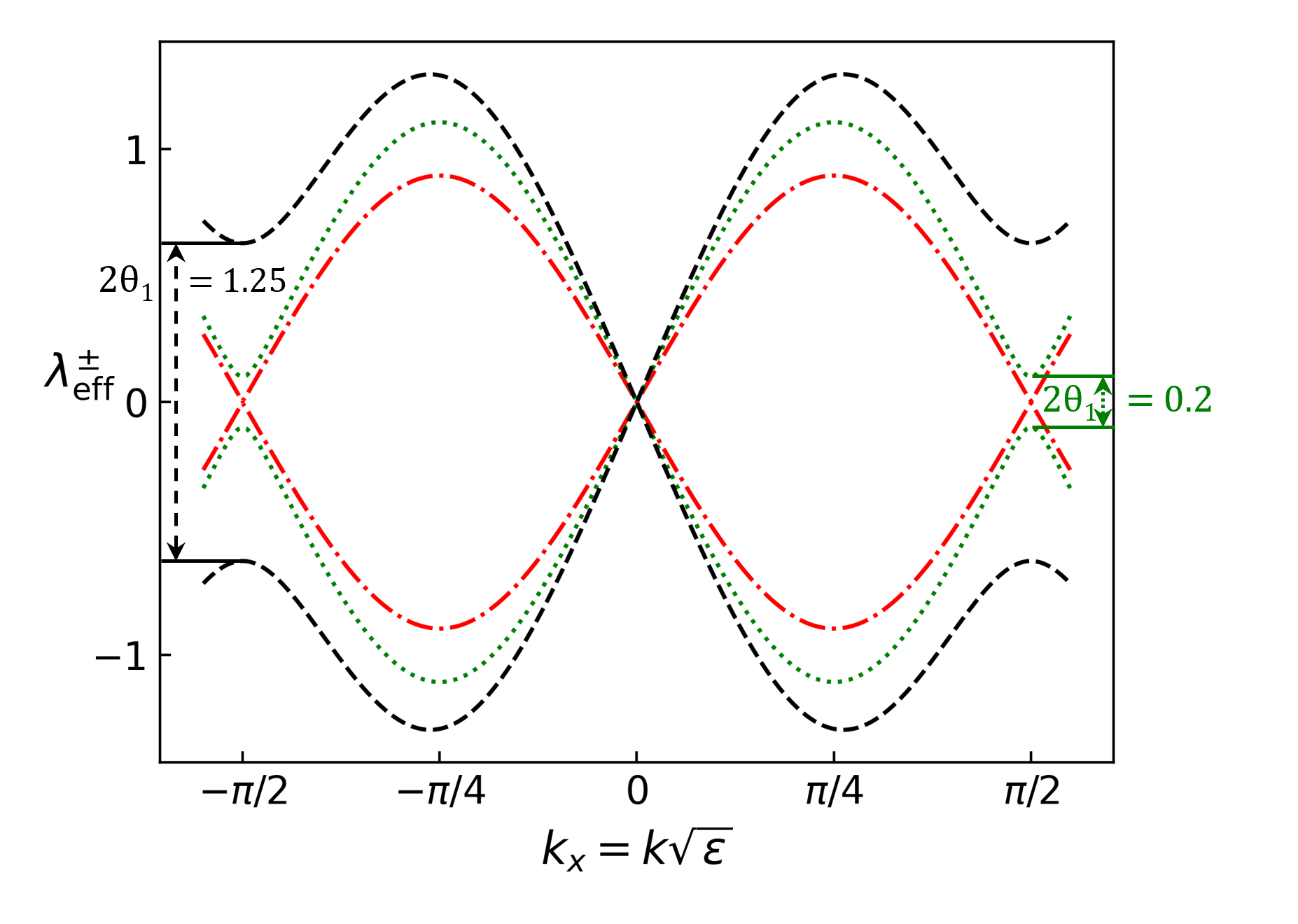}
   \vspace{-10pt}
   \caption{\footnotesize Effective spectrum of the walk as a function of $k$ for different set of parameters $(\alpha, \theta)$. 
   %Solid blue : $(0~,~ \pi/3)$ ; 
   Dash-dot red: $(\pi/3~,~ 0)$ ; Dotted green: $(\pi/4~,~ 0.1)$ ; Dashed black: $(\pi/3~,~ \pi/5)$. All curves are for the step $\varepsilon= 1$. }
\label{fig:spectrum_eff_XZ}
\end{figure}

Following the same methodology of the previous section we can give an exact expression of the effective energy spectrum. Setting again the mass to zero for simplicity, we have: 
\begin{equation}
\begin{aligned}
    d_0=  \frac{A(A+\sin \theta)}{4} &\cdot (\cos 4 k - 1 ) \\ +~ \frac{1-\cos \theta}{2} &\cdot \cos 2 k + \frac{1+\cos \theta}{2} \\ \\
    \text{with: } & A:= \sin 2 \alpha
\end{aligned}
\label{eq:d0_XZ}
\end{equation}
where again the Brillouin Zone is $k \in [-\pi/2, \pi/2]$. We also notice that the spectrum is symmetric in $k$ for all values of the parameters. Moreover, if we focus onto the edges of the BZ, we can remark that the spectrum at $k= \pm \pi/2$ reads as $\lambda_{\text{eff}} = \pm \theta_1$.
In other words, the twist introduces a gap in the symmetric spectrum of exactly of $2 \theta_1$. This term $\theta_1$ acts as an effective mass on the sites lying on the edges of the BZ, as shown in Fig.~(\ref{fig:spectrum_eff_XZ}) for several sets of parameters. Overall, $\theta$ controls the size of the gap and $\alpha$ controls the amplitude of the dominating $\cos 4 k - 1$.

Now let us Taylor expand the spectrum and take the formal limit $\varepsilon \to 0$ in (\eqref{eq:d0_XZ}), we get: 
\begin{equation}
\begin{aligned}
    & d_0 \underrel{\varepsilon \to 0}{\approx} 1 - \frac{1}{2} k^2 (4 \alpha_1 + \theta_1)^2 \varepsilon^2 \\
     & \text{Plus: } \quad  \arccos (1-u)  \underrel{u \underrel{>}{\to} 0}{\approx} \sqrt{2u} \\ \\
    &\text{Hence: } \lambda_{\text{cont}}= \pm  k (2 \alpha_1 + \dfrac{\theta_1}{2})= \pm k \beta
\end{aligned}
\label{eq:spectrum_cont_XZ}
\end{equation}
The overall dynamics is driven by the following Hamiltonian:
% \begin{equation}
% \begin{aligned}
%     \mathcal{H}_c &= \begin{pmatrix*}  -\frac{m}{2} & i \beta k\\
% -i \beta k & + \frac{m}{2}
% \end{pmatrix*}\\
% &= -\dfrac{m}{2} \sigma_z - k \beta \sigma_y
% \end{aligned}
% \end{equation}
\begin{equation}
\begin{aligned}
    \mathcal{H}_c &=&= -\dfrac{m}{2} \sigma_z - k \beta \sigma_y
\end{aligned}
\end{equation}
leading to the couple of PDEs:
\begin{equation}
    i \partial_t \Psi(t,x)= \mathcal{H}_c \Psi
\label{eq:master_eq_XZ}
\end{equation}
The above equations coincide with the well known massive Dirac equation in (1+1) spacetime dimensions. This is not surprising because in this framework the twist angle is of order $\varepsilon$ and the dispersion term is then not present in the final continuous limit. Still both coefficients $\theta_1$ and $\alpha_1$ appears as the prefactor of a first order spatial derivative. 

\section{Conclusion}
\label{sec:conc}

We introduced a new family of quantum walk called, twisted quantum walks. They formally extend the recent plastic quantum walk and bring new remarkable features: they exhibit a richer phenomenology and admits as continuous limit a generalised Dirac equation equipped with a Laplacian. Moreover, quite unexpectedly, we have shown that there exist families of twisted QW which avoid Fermion doubling and seem to be promising as first primitive for building quantum field theories on lattice. \textcolor{black}{Furthermore, we have reason to suspect that this result can be generalised to two and three spatial dimensions. Indeed, a quantum walk higher spatial dimensions can always decompose into one-dimensional quantum walks, one along each spatial dimension. In such a split-step quantum walk one could introduce for each spatial direction a twist, avoiding the doubling. However, this intuition must be rigorously proven and make the subject of future research.}
Moreover, possible future works include certainly a rigorous study in the utmost generality of the TQWs. Indeed, we suspect that the cases discussed in this manuscript are only a fraction of those that can be covered by TQWs. Furthermore, TQWs place a further building block in the understanding of the connection between quantum walkers in discrete and continuous time, that we are aiming to further explore. 

\section{Acknowledgements}
This work is supported by grant ANR-22-CE47-0002-01 from the French National Research Agency and the Amidex fondation and by the PEPR integrated project EPiQ ANR-22-PETQ-0007 part of Plan France 2030. 

\section{Author Contribution Statement}
The authors confirm contribution to the paper as follows: study conception and design~: GDM; data collection:  NJ; analysis and interpretation of results: GDM, NJ; draft manuscript preparation~: GDM and NJ. All authors reviewed the results and approved the final version of the manuscript.

\section{Data Availability Statement}
The numerical code and data used to generate all the figures included in the manuscript can be requested to the corresponding author by email.

\bibliographystyle{ieeetr} 
\bibliography{biblio}

\section{Annex}
\label{sec:annex}

\subsection{The continuous limit}
\label{subsec:contLim}
In this section we detail the continuous limit of the twisted quantum walk equation. As an example, let us consider the homogeneous twist QW in the massless case. In other words, the operator is simply $U_{YY} =G_Y^2$. To simplify the notation in the following we simply drop the $Y$ from the subscript. Let us consider $G^{(0)}$, $G^{(1/2)}$ and $G^{(1)}$ respectively the order $\varepsilon^0$, $\varepsilon^{1/2}$ and $\varepsilon^1$ in the development of the unitary for $\varepsilon \to 0$. The constraints that must be verified in order to admit the continuous limit are: 
\begin{equation}
\begin{aligned}
    \mathbf{(\mathcal{C}_0)} \quad & U_{YY}^{(0)}= (G^{(0)})^2= \mathbb{1}_2 \\
   \mathbf{(\mathcal{C}_{1/2})}\quad & U_{YY}^{(1/2)}= G^{(0)} G^{(1/2)} + G^{(1/2)} G^{(0)}= 0
\end{aligned}
\label{eq:constraints_analytical}
\end{equation}
 
If the above conditions are satisfied, then the leading order is:
\begin{equation}
   U_{YY}^{(1)}= G^{(0)} G^{(1)} + G^{(1)} G^{(0)} + (G^{(1/2)})^2
\label{eq:order_1_analytical}
\end{equation}

We define the following parameters for each matrix of the unitary:
\begin{equation}
\begin{aligned}
    C_{\alpha}= C(\delta_{\alpha}, \theta_{\alpha}, \phi_{\alpha}, \zeta_{\alpha}) &:= e^{i \delta_{\alpha}} R_{z}\left(\zeta_{\alpha}\right) R_{y}\left(\theta_{\alpha}\right) R_{z}\left(\phi_{\alpha}\right) \\
    C_{\beta}= C(\delta_{\beta}, \theta_{\beta}, \phi_{\beta}, \zeta_{\beta} ) &:= e^{i \delta_{\beta}} R_{z}\left(\zeta_{\beta}\right) R_{y}\left(\theta_{\beta}\right) R_{z}\left(\phi_{\beta}\right) \\
    T:= R_{y}\left(\theta\right) 
    \quad &\text{Hence:}\quad   T^{-1}=  R_{y}\left(-\theta\right)\\ \\
    \text{With:}\quad  \theta_{\alpha} &= \theta_{\alpha}^{(0)} + \sqrt{\varepsilon} \theta_{\alpha}^{(1/2)} \\
    \theta_{\beta} &= \theta_{\beta}^{(0)} + \sqrt{\varepsilon} \theta_{\beta}^{(1/2)} 
\end{aligned}
\end{equation}
leaving all other parameters constant. 
Although here we take only a rotation in $y$ for the twist the most general case, that we will not analyse here would be to consider: $R_{z}\left(\zeta\right) R_{y}\left(\theta\right) R_{z}\left(\phi\right) $.

We remember that the shift operator $S_x$ can be expressed simply in Fourier space as:
\begin{equation}
    S_x= e^{i k \sqrt{\varepsilon} \sigma_z}=  R_z(-2 k \sqrt{\varepsilon})
\end{equation}
Then, the constraints $\mathbf{(C_0)}$ and $\mathbf{(C_{1/2})}$ yield conditions on these parameters for the walk to admit a continuous limit, leading to the following theorem:

\begin{theorem}[Homogeneous $Y$ twist]
\label{thm:homo_yy_constant_free}
The Twisted Quantum Walk with a homogeneous $Y$-twist admits a continuous limit if and only if the following set of conditions are met. 

\begin{equation}
\begin{aligned}
    &\text{Constraints on $C_{\alpha}$:} \quad \left\{ 
    \begin{aligned}
    &\phi_{\alpha}= u \pi \text{ or } \theta_{\alpha}^{(0)}= u \pi \\
    &\phi_{\alpha}= \pi/2 +  v \pi \text{ or } \theta_{\alpha}^{(1/2)}= (-1)^{r} \theta_{\beta}^{(1/2)}
    \end{aligned} \right. \\
    &\text{Constraints on $C_{\beta}$:} \quad \left\{ 
    \begin{aligned}
    &\delta_{\beta} = - \delta_{\alpha} + \pi/2 + \ell \pi \\
     &\theta_{\beta}^{(0)} =  (-1)^r  \theta_{\alpha}^{(0)} + \pi + 2 p \pi \\
     &\zeta_{\beta}= \phi_{\alpha} + (r+1) \pi + 2 q \pi \\
     &\phi_{\beta}= -\zeta_{\alpha} + (r+1) \pi + 2 s \pi 
    \end{aligned} \right.
\end{aligned}
\end{equation}
%where $\delta= \delta_{\alpha} + \delta_{\beta}$.

The overall Hamiltonian of the walk in the continuous limit reads:
\begin{equation}
\begin{aligned}
\mathcal{H} &= \sigma_y \Big[ \theta_{\alpha}^{(1/2)} \cos \theta_{\alpha}^{(0)} \sin \phi_{\alpha} k + \\
& \quad \quad \quad \left( \cos \theta_{\alpha}^{(0)} \sin \theta - \cos \theta \sin \theta_{\alpha}^{(0)} \cos \phi_{\alpha} \right) k^2 \Big]
\end{aligned}
\end{equation}
\end{theorem}

\begin{corollary}
The parameters $\phi_{\alpha}$ and $\theta_{\alpha}^{(0)}$ define different families of QW. \underline{Either $\sin \phi_{\alpha}= 0$}, and the $k$ term vanishes, then the Hamiltonian:
\begin{equation}
  \mathcal{H}= \sin \left( \theta \pm \theta_{\alpha}^{(0)} \right) k^2 \sigma_y
\end{equation}
which depends on the sign of $\cos \phi_{\alpha}$. We can notice that this means that even with angle $\theta= 0$, we obtain a $k^2$ term in this case. \underline{Or $\sin \theta_{\alpha}^{(0)}= 0$} in which case, we find a case similar to the example we studied in the main text. The Hamiltonian is:
\begin{equation}
\mathcal{H}= \pm \Big[ \theta_{\alpha}^{(1/2)} \sin \phi_{\alpha} k +  \sin \theta k^2 \Big]  \sigma_y
\end{equation}
In other words, it is the mixed case in between the Dirac case (linear spectrum) and the Schrödinger case (quadratic spectrum) and the angle of the twist $\theta$ pilots the magnitude of the quadratic part.
\end{corollary}

\subsection{Proof of the Theorem}

\subsubsection{Preliminary work and definitions}
Let us start with two results about rotations and commutations. It is straightforward to show that for $\alpha, \beta \in \{x, y, z\} ~;~\alpha \neq \beta$ we have for any angle $\vartheta$: $R_{\alpha}(\vartheta) \sigma_{\beta}= \sigma_{\beta} R_{\alpha}(-\vartheta) $. On the other hand, if the two axis are the same, (i.e., $\alpha= \beta$), then $R_{\alpha}(\vartheta) \sigma_{\beta}= \sigma_{\beta} R_{\alpha}(\vartheta)$.

%We also introduce the following notation: for a given complex 2x2 matrix $M$, we write $\overline{M}:= \sigma_z M \sigma_z$. We note that for any angle $\vartheta$: $\overline{R_z(\vartheta)}= R_z(\vartheta)$ and $\overline{R_y(\vartheta)}= R_y(-\vartheta)$.

In the following, Latin letters after $\ell$ are integer variables used to account for the regularity (symmetry and periodicity) of the constraints.

\

Now, let's express the decomposition of each matrix for $\varepsilon \to 0$.
\begin{equation}
\footnotesize
\begin{aligned}
    S_x &= S_x^{(0)} + S_x^{(1/2)} \sqrt{\varepsilon} + S_x^{(1)} \varepsilon \\
    &= \mathbb{1}_2 + i k \sigma_z \sqrt{\varepsilon} - k^2 /2 \mathbb{1}_1 \varepsilon \\ \\
    C_j &= C_j^{(0)} + C_j^{(1/2)} \sqrt{\varepsilon} + C_j^{(1)} \varepsilon \\
    &= C_j^{(0)} - i \dfrac{\theta_{\alpha}^{(1/2)} }{2} \sigma_y R_z(-2 \zeta_j) C_j^{(0)} \sqrt{\varepsilon} - \dfrac{(\theta_{\alpha}^{(1/2)}) ^2 }{8} C_j^{(0)} \varepsilon \\
    &= C_j^{(0)} - i \dfrac{\theta_{\alpha}^{(1/2)} }{2} C_j^{(0)} R_z(-2 \phi_j) \sigma_y \sqrt{\varepsilon} - \dfrac{(\theta_{\alpha}^{(1/2)}) ^2 }{8} C_j^{(0)} \varepsilon \\
\end{aligned}
\end{equation}

Then let us write the decomposition of $G^{(0)}$. We determine that:
\begin{equation}
\footnotesize
\begin{aligned}
    G^{(0)} &= C_{\beta}^{(0)} \cdot C_{\alpha}^{(0)}
\end{aligned}
\label{eq:thm1_U0}
\end{equation}

\begin{equation}
\footnotesize
\begin{aligned}
     G^{(1/2)} &= T^{-1} S_x^{(1/2)} T^1 C_{\beta}^{(0)} C_{\alpha}^{(0)} + C_{\beta}^{(1/2)} C_{\alpha}^{(0)} + \\
    & \quad C_{\beta}^{(0)} S_x^{(1/2)} C_{\alpha}^{(0)} + C_{\beta}^{(0)} C_{\alpha}^{(1/2)}  \\
    &= -\dfrac{i}{2} \left( \theta_{\alpha}^{(1/2)} J_{\alpha} + \theta_{\beta}^{(1/2)} J_{\beta} \right) \cdot  G^{(0)} \\
    & \quad + i k ~ G^{(0)} \cdot (B + K) \\
\end{aligned}
\label{eq:thm1_U1/2}
\end{equation}

\begin{equation}
\footnotesize
\begin{aligned}
     G^{(1)} &= T^{-1} S_x^{(1)} T^1 C_{\beta}^{(0)} C_{\alpha}^{(0)} + C_{\beta}^{(1)} C_{\alpha}^{(0)} + C_{\beta}^{(0)} S_x^{(1)} C_{\alpha}^{(0)} + C_{\beta}^{(0)} C_{\alpha}^{(1)}  + \\
    & \quad T^{-1} S_x^{(1/2)} T^1 C_{\beta}^{(1/2)} C_{\alpha}^{(0)} + T^{-1} S_x^{(1/2)} T^1 C_{\beta}^{(0)} S_x^{(1/2)} C_{\alpha}^{(0)} + \\
    & \quad T^{-1} S_x^{(1/2)} T^1 C_{\beta}^{(0)} C_{\alpha}^{(1/2)}  + C_{\beta}^{(1/2)} S_x^{(1/2)} C_{\alpha}^{(0)} +  \\
    & \quad C_{\beta}^{(0)} S_x^{(1/2)} C_{\alpha}^{(1/2)} + C_{\beta}^{(1/2)} S_x^{(0)} C_{\alpha}^{(1/2)} \\
    &= -k^2 G^{(0)}B(B+K) \\
    & \quad +  k \Bigg( \dfrac{1}{2} G^{(0)} B G^{(0)} \left( \theta_{\alpha}^{(1/2)} J_{\alpha} + \theta_{\beta}^{(1/2)} J_{\beta} \right)G^{(0)}  + \\
    &\quad \quad \quad \quad i C_{\beta}^{(1/2)} \sigma_z C_{\alpha}^{(0)} + i C_{\beta}^{(0)} \sigma_z  C_{\alpha}^{(1/2)} \Bigg) + \\
   & \quad -\dfrac{1}{8} \left((\theta_{\alpha}^{(1/2)}) ^2+( \theta_{\beta}^{(1/2)}) ^2 \right)  G^{(0)} + C_{\beta}^{(1/2)} C_{\alpha}^{(1/2)}
\end{aligned}
\label{eq:thm1_U1}
\end{equation}

\begin{equation}
\footnotesize
\text{With:    } \left\{
\begin{aligned}
     B &= G^{(0)} \sigma_z T^2 G^{(0)}\\
     K &= \left(C_{\alpha}^{(0)} \right)^{-1} \sigma_z C_{\alpha}^{(0)}=  G^{(0)} C_{\beta}^{(0)}  \sigma_z C_{\alpha}^{(0)} \\
     J_{\alpha} &= R_z(-2 \phi_{\alpha}) \sigma_y\\
     J_{\beta} &= \sigma_y R_z(-2 \zeta_{\beta}) 
\end{aligned} \right.
\end{equation}
It is worth noting that $B^2= K^2= \mathbb{1}_2$. 

Finally, we an express the final 2-step operator of the walk as:
\begin{equation}
\footnotesize
\begin{aligned}
U_{YY} &= \mathbb{1}_2 + \varepsilon ~ U_{YY}^{(1)} \\
\text{With: } \quad U_{YY}^{(1)} &= U_{YY}^{(1), k^2} k^2 + U_{YY}^{(1), k^1} k  + U_{YY}^{(1), 1} 
\end{aligned}
\end{equation}
\begin{equation}
\footnotesize
\text{With: } \quad \left\{ \begin{aligned}
U_{YY}^{(1), k^2} &= \left\{ G^{(1), k^2}, G^{(0)} \right\} + \left( G^{(1/2), k} \right)^2 \\
U_{YY}^{(1), k^1} &= \left\{ G^{(1), k} , G^{(0)} \right\} + \left\{ G^{(1/2), 1} , G^{(1/2), k} \right\} \\
U_{YY}^{(1), ~1} &=  \left\{ G^{(1), 1}, G^{(0)} \right\} + \left( G^{(1/2), 1} \right)^2
\end{aligned}\right.
\label{eq:2step_operator_decomp}
\end{equation}
with $\{ M, N \}= MN + NM$, the anti-commutator.

\

We can now reformulate the conditions $\mathbf{(\mathcal{C}_0)}$ and $\mathbf{(\mathcal{C}_{1/2})}$ for the quantum walk. We find equivalent conditions: 

\begin{equation}
\footnotesize
\begin{aligned}
    \mathbf{(\mathcal{C}_0)} \quad &U_{YY}^{(0)}= (G^{(0)})^2= \mathbb{1}_2 \\
   \mathbf{(\mathcal{C}_{1/2 A})}\quad & \{B + K, G^{(0)}\}= 0 \\
   \mathbf{(\mathcal{C}_{1/2 B})}\quad & \{ \theta_{\alpha}^{(1/2)} J_{\alpha} + \theta_{\beta}^{(1/2)} J_{\beta} , G^{(0)}\}= 0
\end{aligned}
\label{eq:constraints_analytical_proof}
\end{equation}

We can now discuss the implications of these constraints on our parameters. As the product of a complex factor and rotations, the determinant of the zero order unitary is $\det G^{(0)}= e^{2i\delta}$. The condition $\mathbf{(\mathcal{C}_0)}$ imposes that:
\begin{equation}
\footnotesize
    (\det G^{(0)} ) ^2= e^{4i\delta} \stackrel{!}{=} \det \mathbb{1}_2= 1
\end{equation}
which means that we necessarily have $\mathbf{\delta= \ell \dfrac{\pi}{2}} \quad l \in \mathbb{Z}$.

From this, let us distinguish two cases for this proof. 
\subsubsection{First case: $\delta= \ell ' \pi$}
In this case, we have $\det G^{(0)}= 1$. The condition $\mathbf{(\mathcal{C}_0)}$ additionally yields that $G^{(0)}$  is hermitian hence diagonalizable and the spectrum of the unitary is 
\begin{equation}
\footnotesize
    \text{sp}~G^{(0)}= \left\{ \dfrac{\Tr G^{(0)}}{2} \pm \sqrt{ \left( \dfrac{\Tr G^{(0)}}{2} \right) ^2 - \det G^{(0)} } \right\} \stackrel{!}{\subset} \{1, -1\}  
\end{equation}
Given the determinant of $G^{(0)}$, we know that both eigenvalues have the same sign which means that $\mathbf{G^{(0)}= (-1)^m \mathbb{1}_2= \pm \mathbb{1}_2}$. Then we can determine from $\mathbf{(\mathcal{C}_{1/2 A})}$ that $B+K= 0$. And this means that $G^{(1/2), k}= 0$ (Eq. (\eqref{eq:thm1_U1/2}) and $G^{(1), k^2}= 0$ (Eq. (\eqref{eq:thm1_U1}). Hence, from equation (\eqref{eq:2step_operator_decomp}), we know that in this case $U_{YY}^{(1), k^2}= 0 $, in other words, there will not be a $k^2$ term in the Hamiltonian of the continuous limit. The relation $B+K=0$ means that there will constraints liking the angles $\theta$ and those of the operator $C_{\alpha}$, this means that the twist will no longer be free. To avoid this we need to have:
\begin{equation}
\footnotesize
\left\{
\begin{aligned}
\cos \theta  &= - \cos \theta_{\alpha}^{(0)}  \\
 \sin \theta &= - \sin \theta_{\alpha}^{(0)} \cos \phi_{\alpha} \\
 0 &=  \sin \theta_{\alpha}^{(0)} \sin \phi_{\alpha}
\end{aligned}\right.
\end{equation}
or equivalently :
\begin{equation}
\footnotesize
\left\{
\begin{aligned}
\theta_{\alpha}^{(0)}  &= n \pi  \\
\theta &= - \theta_{\alpha}^{(0)} + 2 t \pi \end{aligned}\right. \text{ or } \left\{ \begin{aligned}
\phi_{\alpha} &= n \pi  \\
\theta &= (-1)^n \theta_{\alpha}^{(0)} + \pi + 2 t \pi \end{aligned}\right.
\end{equation}
Still, let us push the analysis a bit further. If we consider condition  $\mathbf{(\mathcal{C}_{1/2 B})}$, it is a reformulation of $C_{\beta}^{(1/2)} C_{\alpha}^{(0)} +
C_{\beta}^{(0)} C_{\alpha}^{(1/2)}= 0$.
We can then use the decomposition of the $C$s to write $\theta_{\alpha}^{(1/2)} R_z(-2\phi_{\alpha})= -\theta_{\beta}^{(1/2)}  R_z(2\zeta_{\beta})$. From this we deduce that $\theta_{\alpha}^{(1/2)}= \pm \theta_{\beta}^{(1/2)}$. If they are equal to 0, then the remaining terms of $U_{YY}^{(1)}$ vanish and the walk is trivial, so we assume they are non zero. 
This means that the rotations must match. When looking at the decomposition of this equation in the basis of the Pauli matrices, we obtain:
\begin{equation}
\footnotesize
    \begin{aligned}
    &\text{If $\theta_{\alpha}^{(1/2)}= \theta_{\beta}^{(1/2)}$: } \quad \sin \phi_{\alpha} + \zeta_{\beta}= 0 \text{ and } \cos \phi_{\alpha} + \zeta_{\beta}= -1 \\
    &\text{If $\theta_{\alpha}^{(1/2)}= - \theta_{\beta}^{(1/2)}$: } \quad \sin \phi_{\alpha} + \zeta_{\beta}= 0 \text{ and } \cos \phi_{\alpha} + \zeta_{\beta}= 1
    \end{aligned}
\end{equation}
If we use the same techniques on the corresponding relationship between $\zeta_{\alpha}$ and $\phi_{\beta}$, we can express: 
\begin{equation}
\footnotesize
    \begin{aligned}
    &\text{If $\theta_{\alpha}^{(1/2)}= \theta_{\beta}^{(1/2)}$: } \quad \sin \zeta_{\alpha} + \phi_{\beta}= 0 \text{ and } \cos \zeta_{\alpha} + \phi_{\beta}= -1 \\
    &\text{If $\theta_{\alpha}^{(1/2)}= - \theta_{\beta}^{(1/2)}$: } \quad \sin \zeta_{\alpha} + \phi_{\beta}= 0 \text{ and } \cos \zeta_{\alpha} + \phi_{\beta}= 1
    \end{aligned}
\end{equation}
We can sum up these relationships by the following statement:
\begin{equation}
\footnotesize
\left\{
\begin{aligned}
 \theta_{\beta}^{(1/2)}  &= (-1)^r  \theta_{\alpha}^{(1/2)} \\
 \zeta_{\beta} &= -\phi_{\alpha} + (r+1) \pi + 2 p \pi \\
 \phi_{\beta} &= -\zeta_{\alpha} + (r+1) \pi + 2 q \pi 
\end{aligned}\right.
\end{equation}
Finally, if we look back at the condition $\mathbf{(\mathcal{C}_{0})}$, it yields that:
\begin{equation}
\footnotesize
\begin{aligned}
G^{(0)} &= (-1)^m \mathbb{1}_2 \\
\Rightarrow R_y(\theta_{\beta}^{(0)} - (-1)^r \theta_{\beta}^{(0)} ) &= (-1)^{m+\ell+p+q+r+1} \mathbb{1}_2
\end{aligned}
\end{equation}
Which simplifies into the last relationship necessary to determine $C_{\beta}$ from $C_{\alpha}$:
\begin{equation}
\footnotesize
\theta_{\beta}^{(0)}= (-1)^r \theta_{\alpha}^{(0)} + (m+\ell+p+q+r+1) 2 \pi + 4 s \pi
\end{equation}
If we sum up all the constraints we obtained, we get:
\begin{equation}
\footnotesize
\begin{aligned}
&\text{Constraints on $C_{\alpha}$ and on the twist:} \\
    &\left\{
\begin{aligned}
\theta_{\alpha}^{(0)}  &= n \pi  \\
\theta &= - \theta_{\alpha}^{(0)} + 2 t \pi \end{aligned}\right. \text{ or } \left\{ \begin{aligned}
\phi_{\alpha} &= n \pi  \\
\theta &= (-1)^n \theta_{\alpha}^{(0)} + \pi + 2 t \pi \end{aligned}\right.
\end{aligned}
\end{equation}
\begin{equation}
\footnotesize
\begin{aligned}
 &\text{Constraints on $C_{\beta}$:} \\
 &\left\{ 
    \begin{aligned}
     \theta_{\beta}^{(1/2)}  &= (-1)^r  \theta_{\alpha}^{(1/2)} \\
 \zeta_{\beta} &= -\phi_{\alpha} + (r+1) \pi + 2 p \pi \\
 \theta_{\beta}^{(0)} &= (-1)^r \theta_{\alpha}^{(0)} + (m+\ell+p+q+r+1) 2 \pi + 4 s \pi \\
 \phi_{\beta} &= -\zeta_{\alpha} + (r+1) \pi + 2 q \pi 
    \end{aligned} \right.
\end{aligned}
\end{equation}
Finally, if we plug in these constraints for the walk, we get:
\begin{equation}
\footnotesize
    \begin{aligned}
    U_{YY}^{(1)} &= k ~ U_{YY}^{(1), k} \\
    &= -2 i k \theta_{\alpha}^{(1/2)}  (\cos \phi_{\alpha} \cos \theta_{\alpha}^{(0)} ,- \sin \phi_{\alpha} \cos \theta_{\alpha}^{(0)} ,  \sin \theta_{\alpha}^{(0)} ) \cdot \vec{\sigma} \\ \\
    &\text{Donc: } \mathcal{H}=  i \dfrac{U_{YY}^{(1)}}{2 \varepsilon}= -\theta_{\alpha}^{(1/2)} k \begin{pmatrix} \cos \phi_{\alpha} \cos \theta_{\alpha}^{(0)} \\ -\sin \phi_{\alpha} \cos \theta_{\alpha}^{(0)} \\ \sin \theta_{\alpha}^{(0)} \end{pmatrix} \cdot \vec{\sigma}
    \end{aligned}
\end{equation}
This means that in this case, the $k^2$ term vanishes, the twist is constrained and the hamiltonian in the continuous limit is Dirac like, with a linear dependence in $k$.

\subsubsection{Second Case: $\delta= \pi/2 + \ell ' \pi$}

In this case, the eigenvalues of $G^{(0)}$ are $1$ and $-1$. This means that $G^{(0)}$ is trace-less and hermitian, so there exists $a, b, c \in \mathbb{C}$, so that $G^{(0)}= a \sigma_x + b \sigma_y + c \sigma_z$. We start with the $\mathbf{(\mathcal{C}_{1/2 A})}$ condition. It yields: 
\begin{equation}
\footnotesize
\begin{aligned}
\{ G^{(0)}, B\} + \{G^{(0)}, K \} &= 0 \\
\{ G^{(0)}, \sigma_z\} \cos \theta - \{ G^{(0)}, \sigma_x\} \sin \theta + \{G^{(0)}, K \} &= 0 \\
\end{aligned}
\end{equation}
As we want to keep the angle of the twist $\theta$ free, we obtain:
\begin{equation}
\footnotesize
\begin{aligned}
\{ G^{(0)}, \sigma_x\}= a &= 0 \\
\{ G^{(0)}, \sigma_z\}= c &= 0 \\
\{G^{(0)}, K \} &= 0
\end{aligned}
\end{equation}
Hence $G^{(0)}= a \sigma_y= \pm \sigma_y$. We also then have $\{\sigma_y, K \}= 0$ which becomes $[R_z(-2\phi_{\alpha}) , R_y(-2\theta_{\alpha}^{(0)}) ]= 0$ and finally the $\sigma_x$ component gives $\sin \phi_{\alpha} \sin \theta_{\alpha}^{(0)}= 0$.  To sum up, the condition $\mathbf{(\mathcal{C}_{1/2 A})}$ gives us:
\begin{equation}
\footnotesize
\mathbf{(\mathcal{C}_{1/2 A})} \Rightarrow \left\{
\begin{aligned}
 G^{(0)}  &= (-1)^n \sigma_y \\
 \sin \phi_{\alpha} \sin \theta_{\alpha}^{(0)} &= 0
\end{aligned}\right.
\label{eq:constr_C1/2A_final}
\end{equation}
If we dig more into the relation $ G^{(0)}=C_{\beta}^{(0)} \cdot C_{\alpha}^{(0)}= (-1)^n \sigma_y$, it gives constraints linking all the parameters of $C_{\beta}^{(0)}$ to the ones of $C_{\alpha}^{(0)}$. More precisely, we get:
\begin{equation}
\footnotesize
C_{\beta}^{(0)}= (-1)^n \sigma_y \left( C_{\alpha}^{(0)} \right)^{-1} \Rightarrow \left\{
\begin{aligned}
 \theta_{\beta}^{(0)}  &= \pi + (-1)^r  \theta_{\alpha}^{(0)} + 2 p \pi \\
 \zeta_{\beta} &= \phi_{\alpha} + (r+1) \pi + 2 q \pi \\
 \phi_{\beta} &= -\zeta_{\alpha} + (r+1) \pi + 2 s \pi 
\end{aligned}\right.
\label{eq:constr_C_beta}
\end{equation}
Now if we look at condition $\mathbf{(\mathcal{C}_{1/2 B})}$, it is quite straightforward to get that $\theta_{\alpha}^{(1/2)} \cos \phi_{\alpha} +\theta_{\beta}^{(1/2)} \cos \zeta_{\beta}= 0$. Using the relationships between $C_{\alpha}$ and $C_{\beta}$ we just obtained, we find that:
\begin{equation}
\footnotesize
\mathbf{(\mathcal{C}_{1/2 B})} \Rightarrow \left( \theta_{\alpha}^{(1/2)} + (-1)^{r+1} \theta_{\beta}^{(1/2)} \right) \cos \phi_{\alpha}= 0
\label{eq:constr_C1/2B_final}
\end{equation}
If we sum up all the constraints we obtained, we get:

\begin{equation}
\footnotesize
    \text{Constraints on $C_{\alpha}$:} \quad \left\{ 
    \begin{aligned}
    &\phi_{\alpha}= u \pi \text{ or } \theta_{\alpha}^{(0)}= u \pi \\
    & \left( \theta_{\alpha}^{(1/2)} + (-1)^{r+1} \theta_{\beta}^{(1/2)} \right) \cos \phi_{\alpha}= 0
    \end{aligned} \right.
\end{equation}
\begin{equation}
\footnotesize
    \text{Constraints on $C_{\beta}$:} \quad \left\{ 
    \begin{aligned}
     &\theta_{\beta}^{(0)} = \pi + (-1)^r  \theta_{\alpha}^{(0)} + 2 p \pi \\
     &\zeta_{\beta}= \phi_{\alpha} + (r+1) \pi + 2 q \pi \\
     &\phi_{\beta}= -\zeta_{\alpha} + (r+1) \pi + 2 s \pi 
    \end{aligned} \right.
\end{equation}
Finally, if we plug in these constraints for the walk, we get:
\begin{equation}
\footnotesize
\begin{aligned}
\mathcal{H} &=  i \dfrac{U_{YY}^{(1)}}{2 \varepsilon} \\
 &= \sigma_y \Big[ \theta_{\alpha}^{(1/2)} \cos \theta_{\alpha}^{(0)} \sin \phi_{\alpha} k + \\
& \quad \quad \quad \left( \cos \theta_{\alpha}^{(0)} \sin \theta - \cos \theta \sin \theta_{\alpha}^{(0)} \cos \phi_{\alpha} \right) k^2 \Big]
\end{aligned}
\end{equation}
\begin{flushright} \qedsymbol \end{flushright}

\subsection{The entanglement entropy in Y-Y case}

Another way to characterise the consequences of introducing a twist into the walker evolution, is through the prism of the entanglement. It is well known that the conditional shift of the walker depending on the coin state induces entanglement between the position and the coin states. This entangling can be measured by the Von Neumann entropy also called the entanglement entropy:
\begin{equation}
    \begin{aligned}
    \mathcal{S} &= - \Tr_C \left[ \rho_{\mathbb{Z}} \log_{2} (\rho_{\mathbb{Z}} ) \right] \\
    &\text{with: } \rho_{\mathbb{Z}}= \Tr_{\mathbb{Z}} \left[~| \Psi \rangle \langle \Psi | ~\right]
    \end{aligned}
\end{equation}

%We can easily process $\mathcal{S}$ at each step of a simulation by diagonalizing $\rho_{\mathbb{Z}}$, the reduced density matrix on the position, which can be written as a $2\times 2$ matrix. 
%Typically, the entropy will start at $0$ and the maximal value reachable by the entropy is $1$ by design for \textit{maximally entangled} states. Here, it means that the spatial degree of freedom and the spin degree of freedom are maximally entangled. Intuitively, for two coupled qubits A and B (two level systems), a \textit{maximally entangled} state is one in which knowing the state of sub-system A gives complete knowledge of sub-system B.
\begin{figure}[hbtp]
\vspace{-5pt}
    \setlength{\belowcaptionskip}{-0pt}
   \includegraphics[width=.5\textwidth]{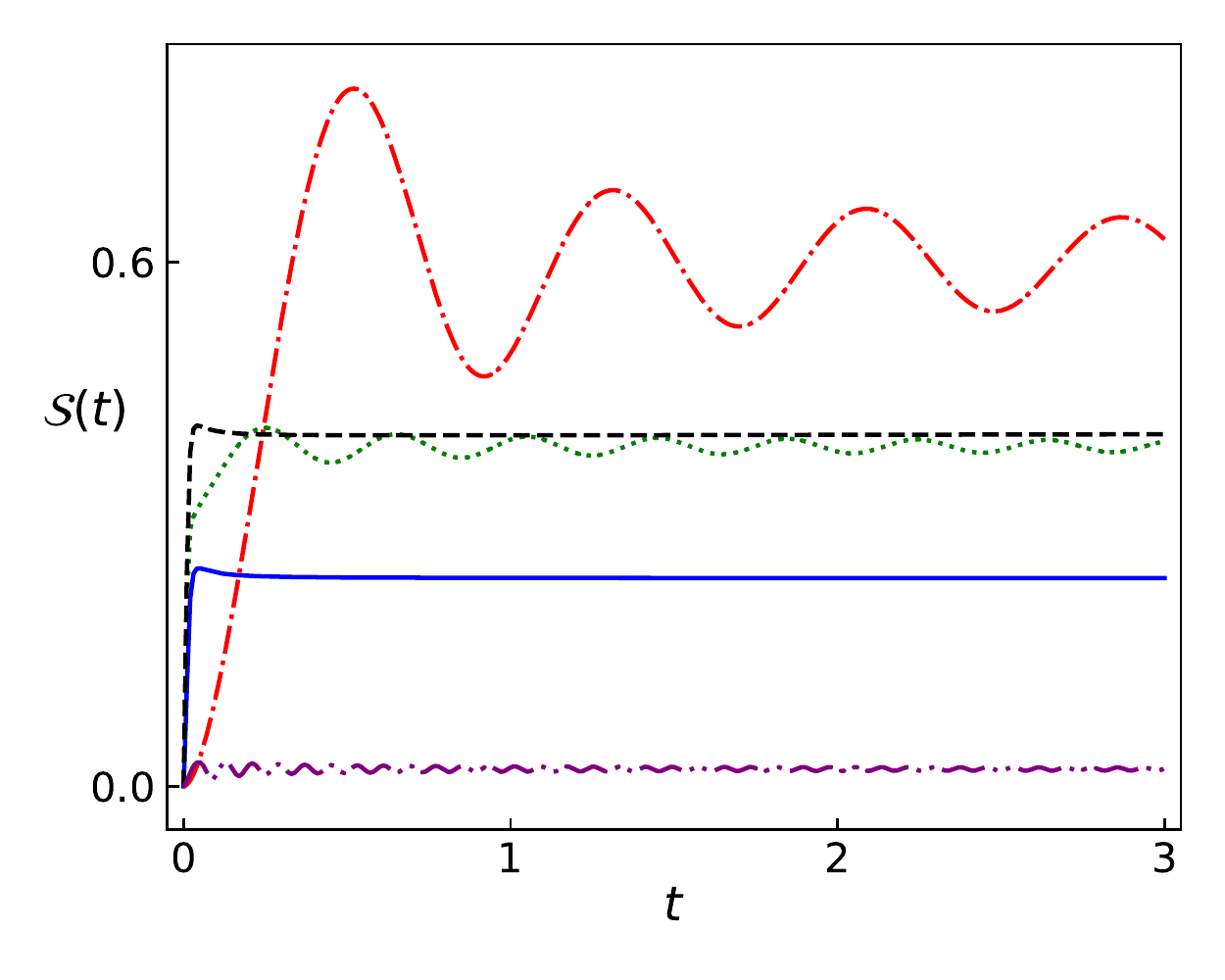}
   \vspace{-15pt}
   \caption{\footnotesize Evolution of $\mathcal{S}$ in time for different set of parameters. The values for the parameters are: 
      Solid blue: $\alpha_1= 0~;~\theta= 2\pi/3$.
      Dash-dot red: $\alpha_1= 0.2~;~\theta= 0$.
      Dotted green: $\alpha_1= 2~;~\theta= \pi/2$.
      Dashed black: $\alpha_1= 0.5~;~\theta= \pi/2$.
      Dash-dot-dot purple: $\alpha_1= 2~;~\theta= \pi$.
     For all data, $\Psi_0 \propto (1,i)^{\text{T}}$, $\varepsilon= 0.01$ and we took $300$ steps. }
    \label{fig:entrop_time_homo}
\end{figure}

Looking at the impact of the various parameters on the entropy, its limit and its behavior, gives insights on the role of each variable. We can draw several conclusions from these observations, analysis we further develop in the Annex~\ref{sec:annex}. First of all, all other things being equal, the eigenvectors $(1, \pm i)$ minimize the limit of the entropy $\mathcal{S}_{\infty}$, which is consistent with our previous description of the walk. These states mainly drift and do not disperse, keeping their initial order. The north and south poles of the Bloch Sphere $(1,0)$ and $(0,1)$ maximize it and always converge to $\mathcal{S}_{\infty}=1$. Overall, we can argue that the presence of the twist \textit{decreases} systematically $\mathcal{S}_{\infty}$ (all other things being equal). There is no difference for $\theta \in 2 \pi \mathbb{Z}$ but for odd multiples of $\pi$, the decrease is maximal. This means that for a twist with an angle $\theta \in \pi + 2  \pi \mathbb{Z}$, the system's degrees of freedom remain much more uncorrelated, especially for initial states close to the eigenvectors. For example, initialized with an eigenvector, the system goes from $\mathcal{S}_{\infty} \approx 0.6$ with no twist to $\mathcal{S}_{\infty} \approx 0$ with twist $\theta= \pi$. We can observe this overall decrease in Fig.~\ref{fig:entrop_time_homo}, the dash-dot red curve (no twist,$\theta= 0$) is above the dashed black and the dotted green curve (twist $\theta= \pi/3$) which are both above the solid blue curve (twist $\theta= 2\pi/3$), which itself is above the dash-dot-dot purple curve (twist $\theta= \pi$).

\subsubsection{More about the entropy in YY case}

\begin{figure}[hbtp]
    \begin{minipage}[t]{.5\textwidth}
    \includegraphics[width=1\textwidth]{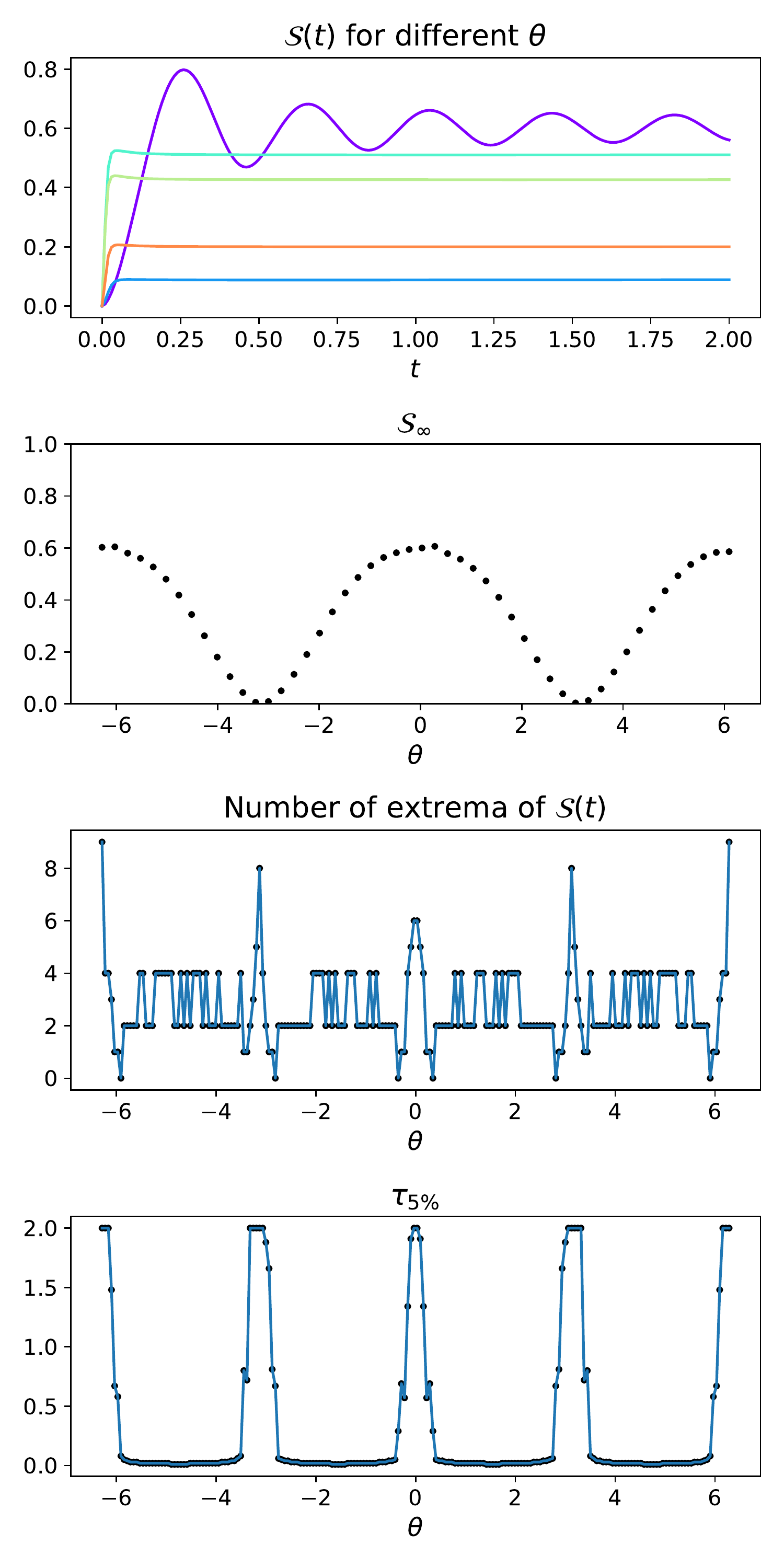}
    \vspace{-20pt}
    \caption{\footnotesize\textbf{a)} $\mathcal{S}$ as a function of time for different values of $\theta \in [-2\pi, 2\pi]$ and with $\alpha_1= 0.4$, and $\Psi_0= \frac{1}{\sqrt{2}}(1,i)^{\text{T}}$. The time-step is $\Delta_t= \epsilon= 0.01$ and the simulation has $200$ steps. Both the limit and the behaviour before equilibrium varies with $\theta$. 
    \textbf{b)} Estimate of $\mathcal{S}_{\infty}$ as a function of $\theta$.
    \textbf{c)} Number of extrema of $\mathcal{S}$ over the run. Here it is mainly between 2 and 4 showing that for most runs, in other words, the oscillations are very slow. We note that for $\theta \sim k \pi$, the oscillations are more numerous.
    \textbf{d)} $\tau_{\text{5\%}}$ as a function of $\theta$. In general it seems to be very low but for $\theta \sim k \pi$ it is particularly high, signifying that when the angle $\theta$ of the twist is multiple of $\pi$, it takes much longer for he entropy to stabilize, it oscillates much more before converging as seen in \textbf{a)} and \textbf{c)}.}
    \label{fig:entrop_var_theta}
    \end{minipage}
 \end{figure}

\begin{figure}[hbtp]
    \begin{minipage}[t]{.5\textwidth}
    \includegraphics[width=1\textwidth]{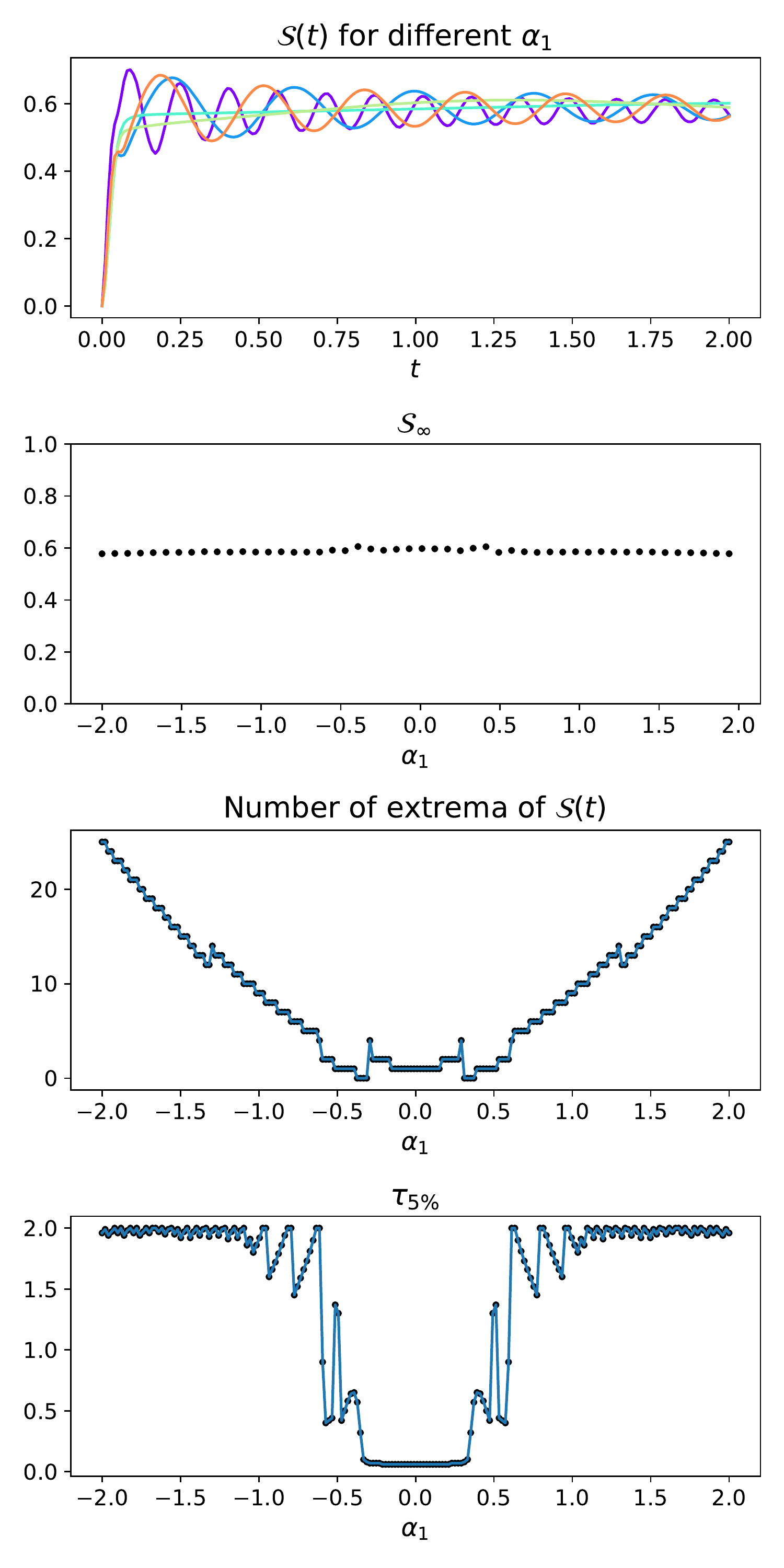}
    \vspace{-20pt}
    \caption{\footnotesize\textbf{a)} $\mathcal{S}$ as a function of time for different values of $\alpha_1 \in [-2, 2]$ and with $\theta= 0.3$, and $\Psi_0= \frac{1}{\sqrt{2}}(1,i)^{\text{T}}$. The time-step is $\Delta_t= \epsilon= 0.01$ and the simulation has $200$ steps. Both the limit and the behaviour before equilibrium varies with $\alpha_1$. 
    \textbf{b)} Estimate of $\mathcal{S}_{\infty}$ as a function of $\alpha_1$. It appears that $\mathcal{S}_{\infty}$ is independent of $\alpha_1$.
    \textbf{c)}It allows us to quantify the oscillatory nature of the transient regime. Here, it appears to increase quasi-linearly in $|\alpha_1|$. In other words, the larger $|\alpha_1|$, the faster the oscillations
    \textbf{d)} $\tau_{\text{5\%}}$ as a function of $\alpha_1$. Similarly to the number of maxima, it appears that for $|\alpha_1| > 0.5$, the oscillations are numerous and with a large amplitude, thus the entropy takes much more time to converge to its limit.}
    \label{fig:entrop_var_alpha}
    \end{minipage}
\end{figure}

\begin{figure}[hbtp]
\begin{minipage}[r]{0.5\textwidth}
    \includegraphics[width=1\textwidth]{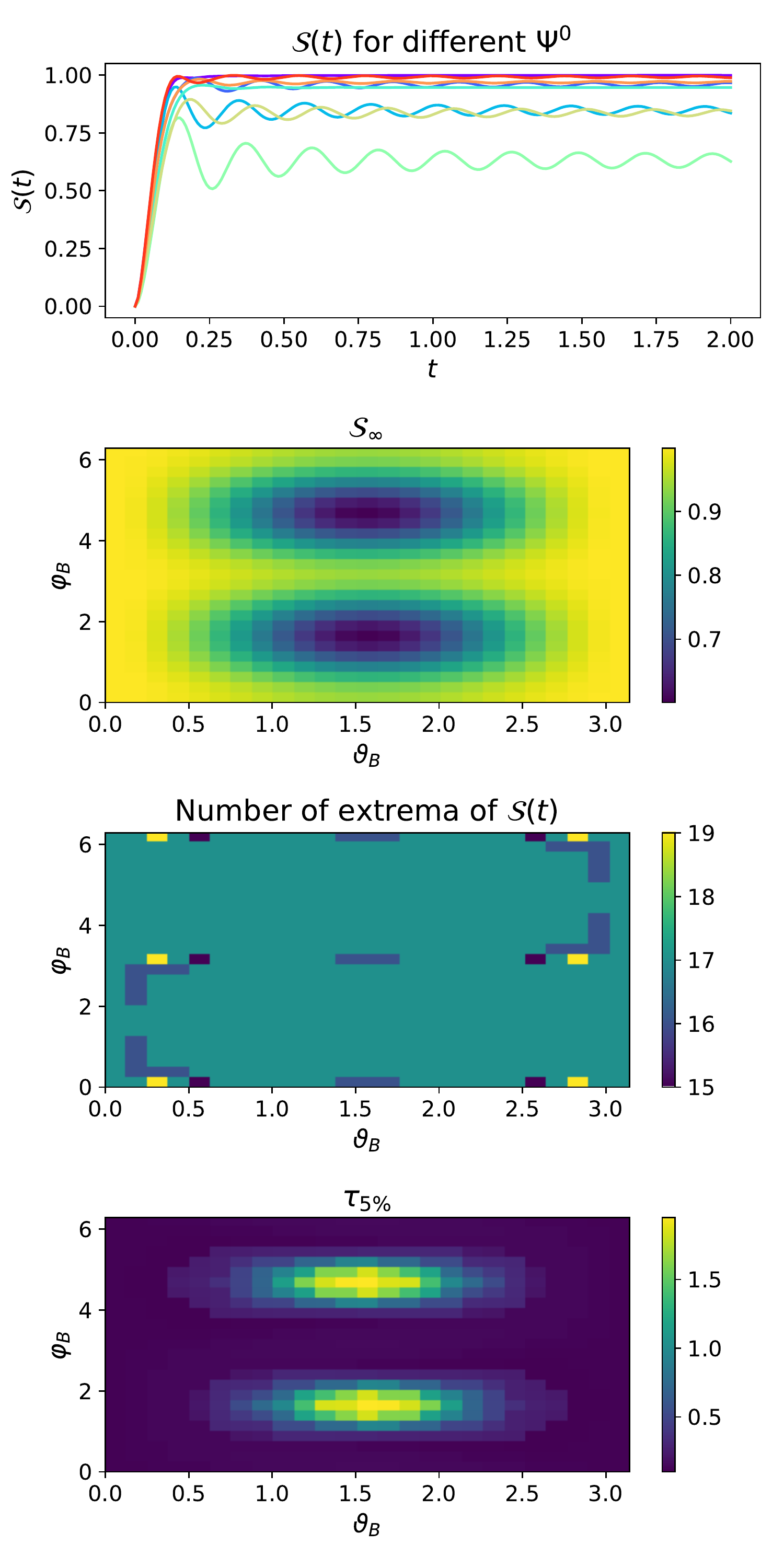}
   \end{minipage}
   \begin{minipage}[r]{0.5\textwidth}
    \caption{\footnotesize\textbf{a)} $\mathcal{S}$ as a function of time for different values of $\Psi^0 (\vartheta_B, \varphi_B)$ ranging over the Bloch Sphere ($\vartheta \in [0, \pi] ~;~\varphi_B \in [0, 2\pi]$) and with $\alpha_1= 0.7$, and $\theta= 0 $. The time-step is $\Delta_t= \epsilon= 0.01$ and the simulation has $200$ steps. Both the limit and the behaviour before equilibrium varies with $\Psi^0$. 
    \textbf{b)} Estimate of $\mathcal{S}_{\infty}$ as a function of $\Psi^0$. The values for $\mathcal{S}_{\infty}$ range from $0.6$ to almost $1$. For example, on the North and South poles ($\Psi_0= (1,0)^{\text{T}}$ or $\Psi_0= (0, 1)^{\text{T}}$), $\mathcal{S}_{\infty}$ is maximal at 1 while for $\vartheta_B= \pi/2,~ \varphi_B= \pi/2$ (i.e.$\Psi_0= \frac{1}{\sqrt{2}}(1, i)^{\text{T}}$ and $\vartheta_B= \pi/2,~ \varphi_B= 3\pi/2$ (i.e.$\Psi_0= \frac{1}{\sqrt{2}}(1,-i)^{\text{T}}$, $\mathcal{S}_{\infty}$ is minimal at 0.6.
    \textbf{c)} Number of extrema of $\mathcal{S}$ over the run. Here it is globally always around 17, in other words, the behaviour of the entropy in its transient regime is always oscillating at about the same frequency.
    \textbf{d)} $\tau_{\text{5\%}}$ as a function of $\Psi^0$. Here it varies widely from almost $0$ to almost $2$. We can see that this is consistent with what we can observe on figure \textbf{a)}.
    }
    \label{fig:entrop_var_CI}
   \end{minipage}
\end{figure}

When looking at time-dependant evolution of $\mathcal{S}$, (Fig.~\ref{fig:entrop_time_homo}), we see that depending, on the couple $(\alpha_1, \theta)$ but also on the initial conditions, the entropy will vary in very different ways, and its asymptotic behaviour even differs. The entropy generally starts with a sharp growth and then either undergoes oscillations around a finite value or converges directly towards a limits. In the general case, the limit is a real number strictly below 1, but some cases seem to lead to an entropy converging to 1, in other words, some parameters seem to lead to a maximally entangled state but it in not the case in general. Let us now characterize the influence of the parameters on the evolution of $\mathcal{S}$. We consider $\alpha_1$, $\theta$ and the initial conditions $\Psi^0$ defined on the Bloch Sphere by the angles $\vartheta_B, \varphi_B$ as:
\begin{equation}
    \ket{\Psi^0}= \begin{pmatrix}  \cos \vartheta_B/2 \\ e^{i\varphi_B} \sin \vartheta_B/2 \end{pmatrix}
\end{equation}
We base this characterization on the experimental simulations we conducted. Overall, the evolution of $\mathcal{S}(t)$ is \textit{symmetric} in $\alpha_1$. It is also symmetric in $\theta$ and seems $2\pi$-periodic. In the initial conditions, it is completely invariant under the $\mathcal{U}(1)$ symmetry. The value for the \textit{limit} of the Entropy $\mathcal{S}_{\infty}$ seems to be independent of $\alpha_1$ (Fig.~\ref{fig:entrop_var_alpha} \textbf{a)}~). As for its dependence in $\theta$, it seems to be oscillating regularly with maxima at $\theta= 2k\pi$ and minima at $\theta= \pi + 2k\pi$ (Fig.~\ref{fig:entrop_var_theta}\textbf{a)}). In the initial conditions, it seems that on the North and South poles ($\Psi_0= (1,0)^{\text{T}}$ or $\Psi_0= (0, 1)^{\text{T}}$), $\mathcal{S}_{\infty}$ is always maximal (for a given pair $(\alpha_1, \theta)$) while for $\vartheta_B= \pi/2,~ \varphi_B= \pi/2$ (i.e.$\Psi_0= \frac{1}{\sqrt{2}}(1, i)^{\text{T}}$ and $\vartheta_B= \pi/2,~ \varphi_B= 3\pi/2$ (i.e. $\Psi_0= \frac{1}{\sqrt{2}}(1,-i)^{\text{T}}$, $\mathcal{S}_{\infty}$ is always minimal (Fig.~\ref{fig:entrop_var_CI} \textbf{a)}). We even note that on the poles, the limit of the entropy seems to be 1 regardless of the two other parameters.

To quantify the \textit{transient regime}, we measured the number of maxima over a simulation as a way to quantify the number of oscillations of the entropy. And we defined $\tau_{\text{5\%}}$ as the time such that $\forall t > \tau_{\text{5\%}} :~ |\mathcal{S}(t) - \mathcal{S}_{\infty} | < 5\%$. It is a way to quantify the speed at which the entropy is converging to its limit. Based on our observations, it seems that the frequency of oscillations of the entropy increases linearly with $|\alpha_1|$. It also seems that the entropy is converges quickly for $\theta \neq k \pi$, reaching $.95~\mathcal{S}_{\infty}$ in $t \approx 0.2$, but much slower with more and broader oscillations for $\theta \sim k \pi$, sometimes taking $10$-time longer to reach $.95\mathcal{S}_{\infty}$. We also note that the convergence towards the limit is much slower for initial conditions close to one of the two sweet spots at $\frac{1}{\sqrt{2}}(1, \pm i)^{\text{T}}$.

 In the continuous limit, using the eigenvalues of the Dirac Hamiltonian expressed in (\eqref{eq:spectrum_cont_homo}), one can derive the entropy of entanglement in this specific case. We find that the eigenvalues of the reduced density matrix $\rho_{\mathbb{Z}}$ are as follows:
\begin{equation}
    \lambda_{\pm}= \dfrac{1}{2} \left[ 1 \pm \sqrt{1+4 |\psi^0_+|^2 |\psi^0_-|^2 \left(\frac{e^{-\frac{4 \alpha_1^2 t^2}{1+ t^2 \sin ^2\theta}}}{\sqrt{1 + t^2 \sin^2 \theta}} - 1\right)} \right]
\end{equation}
From this expression, we can show that in this case, the evolution of the entropy is necessarily monotonous, i.e. it is always increasing towards a finite value only determined by the initial conditions $|\psi^0_+|$ and $|\psi^0_-|$. It is thus qualitatively different from the discrete walk case, in which the entropy is not monotonous, and where its evolution and its asymptotic behaviour is determined only by the set of parameters: \textbf{$\theta$, $\alpha_1$} and the initial spin \textbf{$\Psi_0$}.

\subsection{The entropy in the X-Z case}

The entropy in the X-Z case shows behaviour more complex than if it only depended in the variable $\beta$ and it will deserve a further investigation in the future. Surprisingly for this case the numerical approximation for small $\varepsilon\ll 1$ remains qualitatively different from the analytical continuous limit. Overall, the evolution of $\mathcal{S}(t)$ is \textit{symmetric} in $\beta$. In the initial conditions, it is completely invariant under the $\mathcal{U}(1)$ symmetry. The value for the \textit{limit} of the entropy $\mathcal{S}_{\infty}$ seems to be depend mainly in the initial conditions. It seems that on the North and South poles ($\Psi_0= (1,0)^{\text{T}}$ or $\Psi_0= (0, 1)^{\text{T}}$),$\mathcal{S}_{\infty}$ is always maximal (for a given pair $(\alpha_1, \theta_1 )$) while for $\Psi^0$ slightly off$(1, \pm i)$, $\mathcal{S}_{\infty}$ is always minimal (Fig.~\ref{fig:entrop_var_CI} \textbf{a)}). As for the \textit{transient regime}, it seems that the frequency of oscillations of the entropy increases linearly with $|\beta|$. It also seems that the entropy is converges quickly for $\beta$ close to $0$, reaching $.95\mathcal{S}_{\infty}$ in average in $t \approx 2$ for $|\beta| > 0.5$ but much slower for $|\beta| \ll 1$.

\begin{figure}[hbtp]
    \begin{minipage}[t]{.5\textwidth}
    \includegraphics[width=1\textwidth]{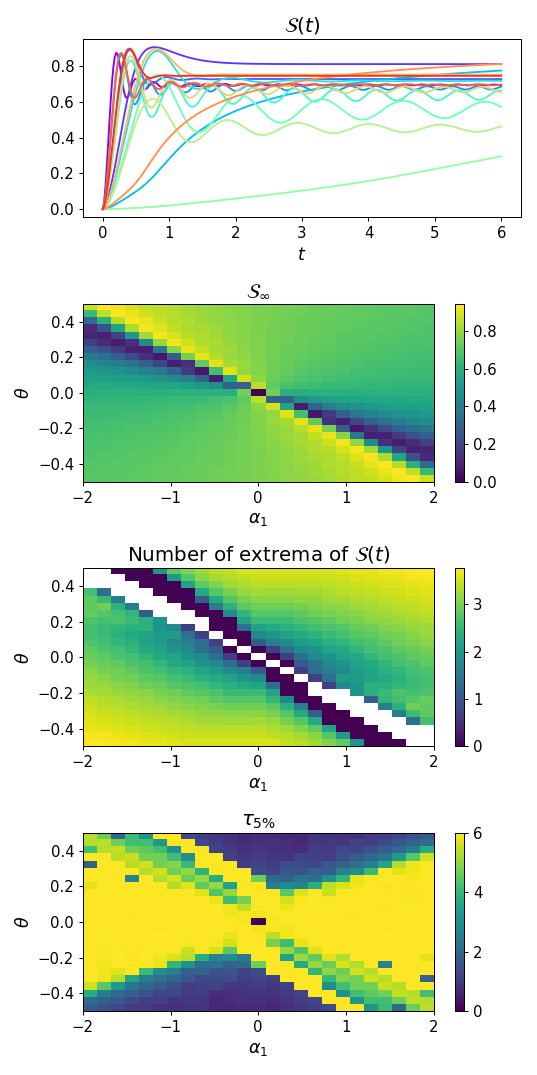}
    \vspace{-20pt}
    \caption{\footnotesize\textbf{a)} $\mathcal{S}$ as a function of time for different values of $\theta_1 \in [-2, 2]$, with $\alpha_1 \in [-0.5, 0.5]$, and $\Psi_0= \frac{1}{\sqrt{2}}(1,i)^{\text{T}}$. The time-step is $\Delta_t= \epsilon= 0.01$ and the simulation has $300$ steps.
    \textbf{b)} Estimate of $\mathcal{S}_{\infty}$ as a function of the parameters.
    \textbf{c)} Number of extrema of $\mathcal{S}$ over the run. 
    \textbf{d)} $\tau_{\text{5\%}}$ as a function of the parameters.}
    \label{fig:entrop_var_param_XZ}
    \end{minipage}
 \end{figure}

\begin{figure}[hbtp]
    \begin{minipage}[t]{.45\textwidth}
    \includegraphics[width=1\textwidth]{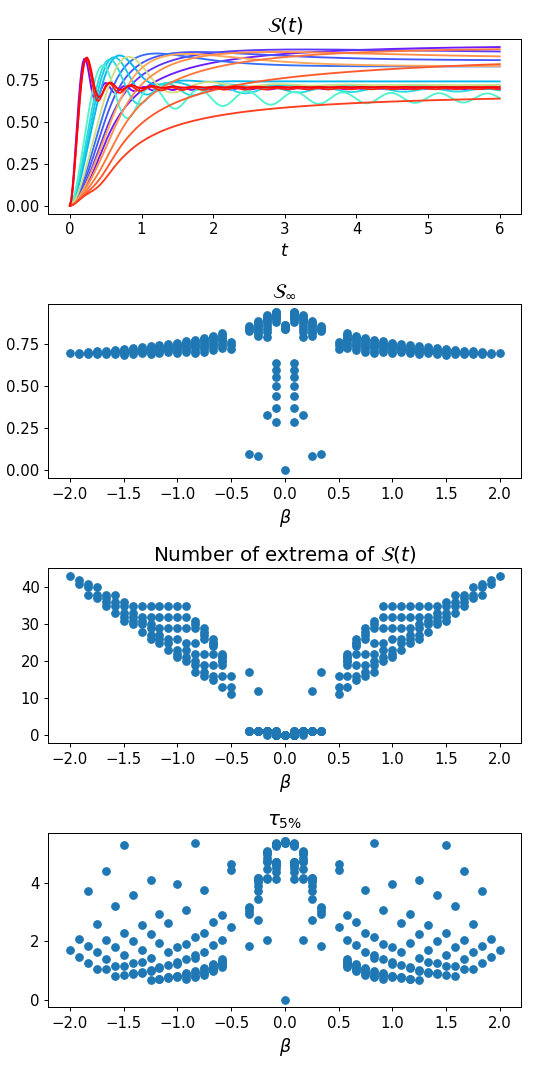}
    \vspace{-20pt}
    \caption{\footnotesize All the data is the one from Fig.~\ref{fig:entrop_var_param_XZ} where we only select the points for which $\tau_{5\%} < 5.5$, i.e. for which the entropy actually gets close to its limit. Here it is presented as function of $\beta= 2\alpha_1 + \theta_1/2$. 
    The overall trend seems to indicate that beta is the main variable controlling the entropy, but even for the limit, different sets of $\alpha_1, \theta_1$ with the same $\beta$ seem to lead to qualitatively different results, especially for small $\beta$.}
    \label{fig:entrop_var_beta_XZ}
    \end{minipage}
\end{figure}

\begin{figure}[hbtp]
\begin{minipage}[r]{0.5\textwidth}
    \includegraphics[width=1\textwidth]{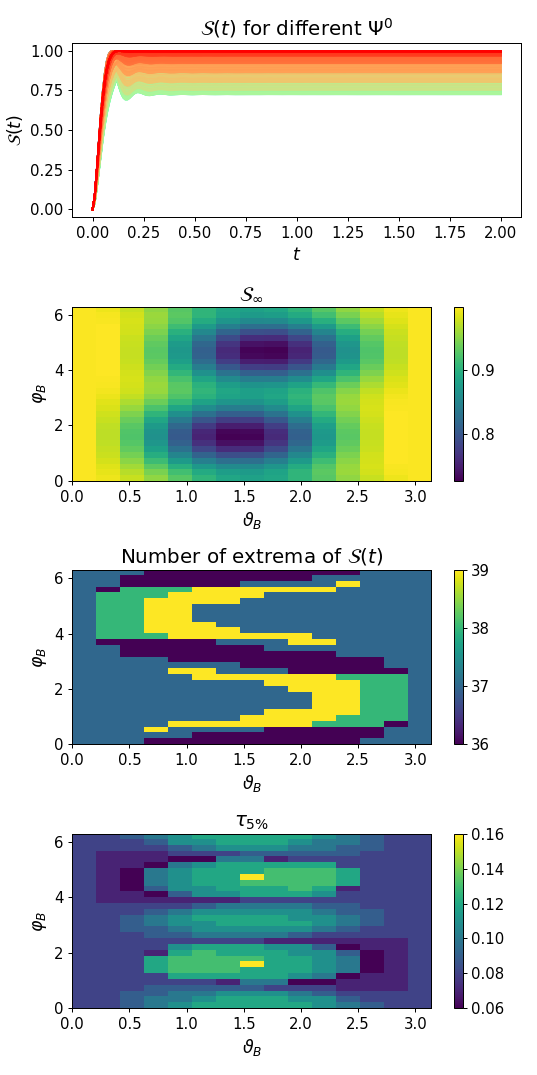}
   \end{minipage}
   \begin{minipage}[r]{0.5\textwidth}
    \caption{\footnotesize\textbf{a)} $\mathcal{S}$ as a function of time for different values of $\Psi^0 (\vartheta_B, \varphi_B)$ ranging over the Bloch Sphere ($\vartheta \in [0, \pi] ~;~\varphi_B \in [0, 2\pi]$) and with $\alpha_1= 0.3$, and $\theta= 1$. The time-step is $\Delta_t= \epsilon= 0.01$ and the simulation has $300$ steps.
    \textbf{b)} Estimate of $\mathcal{S}_{\infty}$ as a function of $\Psi^0$. The values for $\mathcal{S}_{\infty}$ range from $0.7$ to almost $1$, similarly to the Y-Y case. Here as well, on the North and South poles ($\Psi_0= (1,0)^{\text{T}}$ or $\Psi_0= (0, 1)^{\text{T}}$), $\mathcal{S}_{\infty}$ is maximal at 1. However, the minimal points seem slightly of from the previous minima at $(1, \pm i)$. 
    \textbf{c)} Number of extrema of $\mathcal{S}$ over the run. 
    \textbf{d)} $\tau_{\text{5\%}}$ as a function of $\Psi^0$.
    }
    \label{fig:entrop_var_CI_XZ}
   \end{minipage}
\end{figure}
In the continuous limit, using the eigenvalues of the Dirac Hamiltonian expressed in (\eqref{eq:spectrum_cont_XZ}), one can derive the entropy of entanglement in this specific case. We find that the eigenvalues of the reduced density matrix $\rho_{\mathbb{Z}}$ are as follows:
\begin{equation}
\footnotesize
    \lambda_{\pm}= \dfrac{1}{2} \left[ 1 \pm \sqrt{
    \begin{aligned}
    \Im \left[ \psi_+^0 \overline{\psi_-^0} \right]^2 + e^{-\frac{t^2 \beta^2}{\sigma^2}}& \boldsymbol{\cdot} \\ 
\bigg( 1 + 4|\psi_+^0|^4 - 4& |\psi_+^0|^2 + 4\Re \left[ \psi_+^0 \overline{\psi_-^0} \right]^2 \bigg)
\end{aligned} 
\quad} \right]
\end{equation}
From this expression, we can show that in this case, the evolution of the entropy is necessarily monotonous, i.e. it is always increasing towards a finite value only determined by the initial conditions $|\psi^0_+|$ and $|\psi^0_-|$. It is thus qualitatively different from the discrete walk case, in which the entropy is not monotonous, and where its evolution and its asymptotic behaviour is determined only by the set of parameters: \textbf{$\beta$} and the initial spin \textbf{$\Psi_0$}. Also, this formula predicts that in the continuous limit and for the eigenstates in this limit, $(1, \pm i)$, the entropy is consistently $0$ when $\beta \neq 0$. This means that slight discrepancy we observe in Fig.~\ref{fig:entrop_var_CI_XZ} is due to the step being $\varepsilon \neq 0$.

\end{document}